\documentclass[epj]{svjour}
\usepackage{graphics}
\usepackage{color}
\usepackage{comment}
\usepackage{amsmath,amssymb}
\usepackage[OT2,T1]{fontenc}
\usepackage{acronym}
\usepackage{aas_macros}

\newcommand{\talys}{{\sf{TALYS}}}

\begin{document}

\title{PANDORA project: photo-nuclear reactions below $A=60$}
\titlerunning{PANDORA project: photo-nuclear reactions below $A=60$}
\authorrunning{A. Tamii, L. Pellegri, P.-A. S\"{o}derstr\"{o}m, {\it et al.}}

\author{
A.~Tamii\inst{1,2,3,*}, 
L.~Pellegri\inst{4,5}, 
P.-A.~S\"{o}derstr\"{o}m\inst{6}, 
D.~Allard\inst{7}, 
S.~Goriely\inst{8}, 
T.~Inakura\inst{9}, 
E.~Khan\inst{10}, 
E.~Kido\inst{11}, 
M.~Kimura\inst{12,13,11}, 
E.~Litvinova\inst{14}, 
S.~Nagataki\inst{11}, 
P.~von~Neumann-Cosel\inst{15}, 
N.~Pietralla\inst{15}, 
N.~Shimizu\inst{16}, 
N.~Tsoneva\inst{6}, 
Y.~Utsuno\inst{17}, 
S.~Adachi\inst{18}, 
P.~Adsley\inst{19,20}, 
A.~Bahini\inst{5}, 
D.~Balabanski\inst{6}, 
B.~Baret\inst{10}, 
J.A.C.~Bekker\inst{4,5}, 
S.D.~Binda\inst{4,5}, 
E.~Boicu\inst{6,21}, 
A.~Bracco\inst{22,23}, 
I.~Brandherm\inst{15}, 
M.~Brezeanu\inst{6,21}, 
J.W.~Brummer\inst{5}, 
F.~Camera\inst{22,23}, 
F.C.L.~Crespi\inst{22,23}, 
R.~Dalal\inst{24}, 
L.M.~Donaldson\inst{5}, 
Y.~Fujikawa\inst{25}, 
T.~Furuno\inst{3}, 
H.~Haoning\inst{14}, 
Y.~Honda\inst{3}, 
A.~Gavrilescu\inst{6,26}, 
A.~Inoue\inst{1}, 
J.~Isaak\inst{15}, 
H.~Jivan\inst{4,5}, 
P.M.~Jones\inst{5}, 
S.~Jongile\inst{5}, 
O.~Just\inst{11,27}, 
T.~Kawabata\inst{3}, 
T.~Khumalo\inst{4,5}, 
J.~Kiener\inst{10}, 
J.~Kleemann\inst{15}, 
N.~Kobayashi\inst{1}, 
Y.~Koshio\inst{28}, 
A.~Ku\c{s}o\u{g}lu\inst{6,29}, 
K.C.W.~Li\inst{30}, 
K.L.~Malatji\inst{5}, 
R.E.~Molaeng\inst{4,5}, 
H.~Motoki\inst{12}, 
M.~Murata\inst{1}, 
A.A.~Netshiya\inst{4,5,31}, 
R.~Neveling\inst{5}, 
R.~Niina\inst{1}, 
S.~Okamoto\inst{25}, 
S.~Ota\inst{1}, 
O.~Papst\inst{15}, 
E.~Parizot\inst{10}, 
T.~Petruse\inst{6}, 
M.S.~Reen\inst{32}, 
P.~Ring\inst{33}, 
K.~Sakanashi\inst{3}, 
E.~Sideras-Haddad\inst{4}, 
S.~Siem\inst{30}, 
M.~Spall\inst{15}, 
T.~Suda\inst{34}, 
T.~Sudo\inst{1}, 
Y.~Taniguchi\inst{35}, 
V.~Tatischeff\inst{10}, 
H.~Utsunomiya\inst{36,37}, 
H.~Wang\inst{36,38,39}, 
V.~Werner\inst{15}, 
H.~Wibowo\inst{40}, 
M.~Wiedeking\inst{4,5}, 
O.~Wieland\inst{22,23}, 
Y.~Xu\inst{6} 
and
Z.H.~Yang\inst{41} 
(PANDORA Collaboration)
}
\institute{\tiny
Research Center for Nuclear Physics (RCNP), Osaka University, Ibaraki 567-0047, Japan
\and
Institute for Radiation Sciences (IRS), Osaka University, Toyonaka, Osaka 560-0043, Japan
\and
Department of Physics, Osaka University, Toyonaka, Osaka 560-0043, Japan
\and
School of Physics, University of the Witwatersrand, Johannesburg 2050, South Africa
\and
iThemba Laboratory for Accelerator Based Sciences, Somerset West 7129, South Africa
\and
Extreme Light Infrastructure-Nuclear Physics (ELI-NP)/Horia Hulubei National Institute for Physics and Nuclear Engineering (IFIN-HH), Str. Reactorului 30, Bucharest-M\u{a}gurele 077125, Romania
\and
Laboratoire Astroparticule et Cosmologie, Universit\'{e} Paris Cit\'{e}, CNRS, F-75013 Paris, France
\and
Institut d'Astronomie et d'Astrophysique, Universit\'e Libre de Bruxelles, Campus de la Plaine CP 226, 1050 Brussels, Belgium
\and
Tokyo Institute of Technology, 2-12-1 Ookayama, Meguro, Tokyo 152-8550, Japan
\and
IJCLab, Universit\'e Paris-Saclay, CNRS/IN2P3, 91405 Orsay Cedex, France
\and
RIKEN, 2-1 Hirosawa, Wako, Saitama, 351-0198, Japan
\and
Department of Physics, Hokkaido University, Sapporo 060-0810
\and
Japan and Nuclear Reaction Data Centre, Faculty of Science, Hokkaido University, Sapporo 060-0810, Japan
\and
Department of Physics, Western Michigan University, Kalamazoo, Michigan 49008, USA
\and
Institut f\"{u}r Kernphysik, Technische Universit\"{a}t Darmstadt, 64289 Darmstadt, Germany
\and
Center for Computational Sciences, University of Tsukuba, Tsukuba 305-8577, Japan
\and
Advanced Science Research Center, Japan Atomic Energy Agency, Tokai, Ibaraki 319-1195, Japan
\and
Cyclotron and Radioisotope Center, Tohoku University, 6-3 Aoba Aramaki, Aoba, Sendai, Miyagi 980-8578, Japan
\and
Department of Physics \& Astronomy, Texas A\&M University, College Station, 77843-4242, USA
\and
Cyclotron Institute, Texas A\&M University, College Station, 77843-3636, USA
\and
University of Bucharest, Atomistilor 405, 077125 Bucharest-M\u{a}gurele, Romania
\and
Dipartimento di Fisica dell'Universit\`{a} degli Studi di Milano, I-20133 Milano, Italy
\and
INFN, Sezione di Milano, I-20133 Milano, Italy
\and
Guru Jambheshwar University of Science and Technology, Hisar, India-125001
\and
Department of Physics, Kyoto University, Kitashirakawa Oiwake-Cho, 606-8502 Kyoto, Japan
\and
Politehnica University of Bucharest, Bucharest, Romania
\and
GSI Helmholtzzentrum f\"{u}r Schwerionenforschung, Planckstrasse 1, D-64291 Darmstadt, Germany
\and
Department of Physics, Okayama University, 700-8530 Okayama, Japan
\and
Department of Physics, Faculty of Science, Istanbul University, Vezneciler/Fatih, 34134, Istanbul, Turkey
\and
Department of Physics, University of Oslo, P.O. Box 1048, Blindern, N-0316 Oslo, Norway
\and
Department of Chemical and Physical sciences, Walter Sisulu University, Mthatha 5100, South Africa
\and
Department of Physics, Akal University, Talwandi Sabo, Bathinda, Punjab 151302, India
\and
Physics Department, TU Munich, D-85748 Garching, Germany
\and
Research Center for Electron-Photon Science, Tohoku University, 982-0826, Sendai, Japan
\and
Department of Information Engineering, National Institute of Technology, Kagawa college, Takamatsu 761-8058, Japan
\and
Shanghai Advanced Research Institute, Chinese Academy of Sciences, Shanghai 201210, China
\and
Department of Physics, Konan University, Kobe 658-8501,Japan
\and
Shanghai Institute of Applied Physics, Chinese Academy of Sciences, Shanghai 201800, China
\and
University of Chinese Academy of Science, Beijing 100049,China
\and
Department of Physics, University of York, Heslington, York YO10 5DD, United Kingdom
\and
School of Physics and State Key Laboratory of Nuclear Physics and Technology, Peking University, Beijing 100871, China
}

\date{
}

%
\abstract{
Photo-nuclear reactions of light nuclei below a mass of $A=60$ are studied experimentally and theoretically by the PANDORA (Photo-Absorption of Nuclei and Decay Observation for Reactions in Astrophysics) project.
Two experimental methods, virtual-photon excitation by proton scattering and real-photo absorption by a high-brilliance $\gamma$-ray beam produced by laser Compton scattering, will be applied to measure the photo-absorption cross sections and the decay branching ratio of each decay channel as a function of the photon energy.
Several nuclear models, {\em e.g.} anti-symmetrized molecular dynamics, mean-field type models, a large-scale shell model, and {\em ab initio} models, will be employed to predict the photo-nuclear reactions.
The uncertainty in the model predictions will be evaluated from the discrepancies between the model predictions and the experimental data. The data and the predictions will be implemented in a general reaction calculation code \talys .
The results will be applied to the simulation of the photo-disintegration process of ultra-high-energy cosmic rays in inter-galactic propagation.
\PACS{
{25.20.-x}{Photonuclear reactions}
\and
{25.70.De}{Coulomb excitation}
\and
{24.30.Cz}{Giant resonances}
\and
{96.40.-z}{Cosmic rays}
}
} 
\maketitle

\section{Introduction}

\label{sec:introduction}



Photo-nuclear reaction data provide fundamental information on different aspects of nuclear structure, collective excitations, and the response of nuclei to an external electric-dipole ($E1$) field~\cite{bracco2019,Zilges2022}. The \ac{IVGDR} dominates the $E1$ response of nuclei and is a widely-known example of nuclear collective excitation, described as a relative dipole oscillation between neutrons and protons. 
The isovector properties of the nuclear matter are constrained by systematic analyses of the $E1$ response of nuclei, such as the mean excitation energy of the \ac{IVGDR}, strength concentration around the neutron threshold, often called \ac{PDR}~\cite{SavranAumannZilges2013,Roca-Maza2018}, and the static electric-dipole polarizability~\cite{Neumann-Cosel2019electric,TamiiPoltoratskaNeumannEtAl2011,Roca-MazaVinasCentellesEtAl2015} that corresponds to the inversely-energy-weighted sum-rule value of the $E1$ strength distribution. Photo-nuclear reaction data and predictions are crucial for understanding various astrophysical processes, such as the $r$-process nucleosynthesis and the intergalactic propagation of \ac{UHECR}. They are also crucial for applications such as radiation shield design, aspects of non-destructive testing, $\gamma$-ray imaging, isotope production, nuclear medicine, and our understanding of biological responses to radiation. 

\begin{figure}
\begin{center}
\resizebox{0.5\textwidth}{!}{%
  \includegraphics{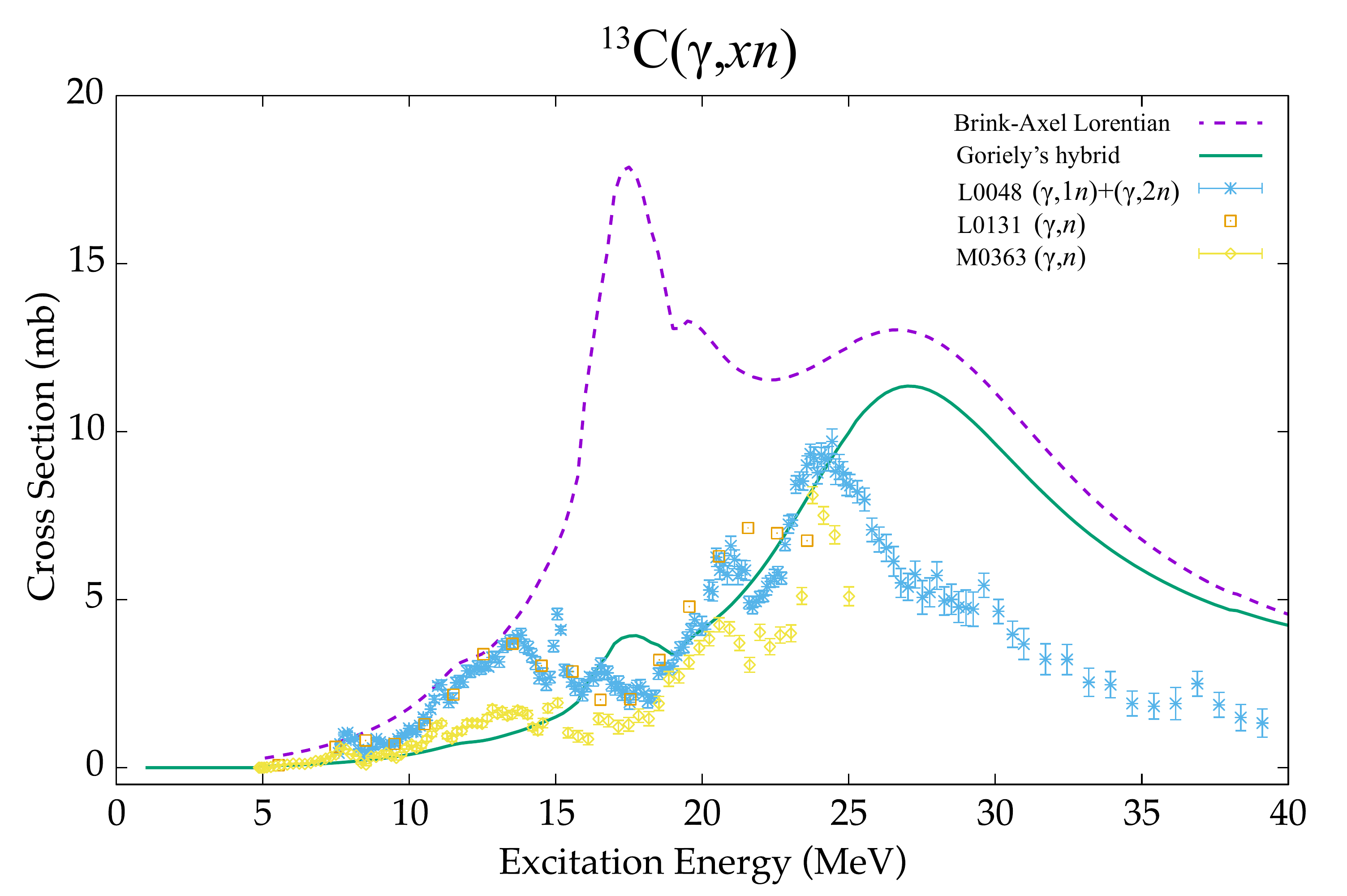}
}
\caption{
Photo-neutron cross sections, $^{13}{\rm C}(\gamma,xn)$, as a function of the excitation energy (photon-energy). The theoretical cross sections are calculated by the reaction code \talys\ using two models, the Brink-Axel Lorentian (dashed line) and the Goriely's hybrid (solid line). They are plotted in comparison with available experimental data taken from the EXFOR database~\cite{EXFOR}. 
The L0131~\cite{maeda2006isovector} data (cross marks) were measured with a tagged-photon beam, M0363~\cite{koch1976photoneutron} (diamond) with a Bremsstrahlung photon-beam, and L0048~\cite{jury1979photoneutron} (square) by the positron-annihilation in-flight method.
L0131 and M0363 (diamond) are one neutron detection data, while L0048~\cite{jury1979photoneutron} is the integrated cross-section of the $(\gamma,n)$ and $(\gamma,2n)$ channels. 
The two neutron separation energy in $^{13}{\rm C}$ is $S_{nn}=23.7$ MeV. 
}
\label{fig:13c}
\end{center}
\end{figure}
The $E1$ response of nuclei has been widely studied in the mass region above $A\sim90$, where nuclei show bulk and smooth properties of nuclear matter with less dependence on the shell structure than in the lower mass region. Experimentally the $(\gamma,xn)$ reaction cross sections, corresponding to the neutron emission channels, are often treated as a good approximation of the total photo-absorption cross-section.

The approximation is not justified in the mass region below $A=60$ due to a comparable or even larger branching ratio to the proton decay channel than the neutron. 
The $\alpha$-decay branching ratio is also not negligible in light nuclei.
Reliable prediction of the photo-nuclear reactions in the lower mass region is more complicated than for heavier nuclei due to the bigger impact of effects such as nuclear shell structure, $\alpha$-clustering, isospin-selection rule, and isospin-mixing in the $\alpha$-decay processes, nuclear deformation, $n$-$p$ pairing, and pre-equilibrium decay processes. Additionally, the \ac{IVGDR} distribution is often fragmented into several pieces in contrast to the case of heavier nuclei.
The experimental data are scarce, especially for charged particle decay channels and the available data are often mutually inconsistent.
The parameters of the theoretical models are usually optimized for heavier nuclei. When used to predict the behavior of lighter nuclei, the predictions are not good enough in most cases.
The neutron decay channel of the photo-nuclear reaction on $^{13}{\rm C}$ is shown in Fig.~\ref{fig:13c} as a representative example.

The \ac{PANDORA} project has been initiated to study the photo-nuclear reaction of light nuclei.
The project relies on interdisciplinary research across the following fields: experimental and theoretical nuclear physics and particle astrophysics. 
A precise and systematic data set consisting of the photo-absorption cross sections, and the branching ratio of each decay channel will be established through the experimental nuclear physics component of the project.
Two complementary methods will be employed to obtain the experimental data: virtual-photon excitation by relativistic proton scattering at very forward angles and real-photon excitation by utilizing a quasi-monoenergetic $\gamma$-ray beam. The details are described in Sec.~\ref{sec:methods}.

In the theoretical nuclear physics part of the project, models will be developed to predict the photo-absorption cross sections and decay branching ratios.
The employed models will include \ac{AMD} \cite{kimura2016,taniguchi2004}, mean-field type calculations - \ac{RNFT} \cite{litvinova2008,litvinova2010,robin2016}, \ac{RPA} with density functional approach \cite{inakura2011} and \ac{QPM} + \ac{EDF} \cite{Tso04,Tso16}-, large-scale shell-model calculations \cite{utsuno2015}, and {\em ab initio} type calculations.
Predictions will be evaluated with the experimental data to estimate the model uncertainties and further developments. 

The obtained experimental data and the model predictions will be implemented in a general reaction calculation code, \talys\, to be made available for applications in various fields.
An important outcome of the \ac{PANDORA} project will be a better characterization of the photo-interactions of \ac{UHECR}, which are an essential ingredient of the theoretical modeling of their extra-galactic propagation and of their acceleration in astrophysical sources. This phenomenon will be briefly described in Sec.~\ref{sec:UHECR}.


This article is organized as follows. 
Experimental methods for measuring the photo-absorption cross sections and the decay branching ratios are described in Sec.~\ref{sec:experiment}.
Theoretical models are described in Sec.~\ref{sec:theory}.
The background of the \ac{UHECR} physics and the application of the photo-nuclear reaction information to the extra-galactic propagation of \ac{UHECR}s are introduced in Sec.~\ref{sec:UHECR}. 
Other potential applications are briefly shown in Sec.~\ref{sec:applications} followed by the summary and outlook in Sec.~\ref{sec:summary}.

\section{Experimental Nuclear Physics} 
\label{sec:experiment}
An overview of the experimental methods is described in Sec.~\ref{sec:methods}. Then the three main experimental facilities, RCNP, iThemba LABS, and ELI-NP of the PANDORA project, are introduced followed by the other related facilities.
\subsection{Methods}\label{sec:methods}


The experimental nuclear physics part of the \ac{PANDORA} project aims to systematically measure the photo-absorption cross sections and the $p$, $n$, $\alpha$, and $\gamma$ decay branching ratios for stable nuclei in the mass region below $A=60$.
Two modern experimental methods will be employed.
One is the virtual photon excitation by relativistic proton scattering applicable at the \ac{RCNP} in Osaka University, Japan, and at iThemba LABS, South Africa ~\cite{Neumann-Cosel2019electric,tam09,nev11}.
The other method is the real photon excitation by high-intensity quasi-monoenergetic $\gamma$ beams produced by \ac{LCS}
at \ac{ELI-NP}, Romania.
The combination of the complementary facilities is essential to obtain a high-quality systematic data set.
The \ac{ELI-NP} facility~\cite{Gales2016} is under construction as a next-generation high-brilliance high-resolution \ac{LCS} facility following the successful operation of the \ac{LCS} facilities in Japan~\cite{toyokawa1999,kawano2020} and in US~\cite{pietralla2002}.

To establish the consistency among the data measured at the different facilities $^{27}{\rm Al}$ was chosen as the reference nucleus for each experimental campaign due to ease of fabrication and treatment, natural mono-isotopic abundance, availability of the $(\gamma,abs)$ data~\cite{ahr75}, and prediction of reasonably large branching ratios for each of the $p$, $n$ and $\alpha$ decay channels with relatively low threshold energies.
All three facilities can identify the nuclear excitation energy with a high resolution of 100 keV or better. This excellent energy resolution is beneficial to establish mutual data consistency between the three facilities for discrete states, broad resonances, and related fine structures.

Almost all the stable nuclei below a mass of $A=60$ are involved in the photo-disintegration path of the \ac{UHECR} nuclei in the intergalactic propagation (see Sec.~\ref{sec:UHECR}). 
Since it is impractical to measure the photo-nuclear reactions for all the stable nuclei involved, a limited set of nuclei will be measured to allow the bench-marking of the nuclear model predictions.
A selection of candidate nuclei is listed below.
\begin{itemize}
    \item reference target: $^{27}{\rm Al}$
    \item first cases and importance of $\alpha$-decay: $^{12}{\rm C}$ and $^{16}{\rm O}$
    \item light nuclei: $^{6}{\rm Li}$, $^{7}{\rm Li}$, $^{9}{\rm Be}$, and $^{10,11}{\rm B}$
    \item $N=Z$ nuclei, $\alpha$-cluster effect and deformation: $^{24}{\rm Mg}$, $^{28}{\rm Si}$, $^{32}{\rm S}$ and $^{40}{\rm Ca}$
    \item isospin selection-rule in $\alpha$-decay: isotope pairs of $^{10,11}{\rm B}$, $^{12,13}{\rm C}$, $^{16,18}{\rm O}$ and $^{24,26}{\rm Mg}$.  
    \item $N>Z$ nuclei and multi-neutron emission: $^{13}{\rm C}$, $^{18}{\rm O}$, $^{26}{\rm Mg}$, $^{48}{\rm Ca}$ and $^{56}{\rm Fe}$
    \item $A$-odd and odd-odd nuclei: $^{7}{\rm Li}$, $^{9}{\rm Be}$, $^{10,11}{\rm B}$, $^{13}{\rm C}$, $^{14}{\rm N}$
\end{itemize}
Self-supporting pure targets or chemical compounds are used with high isotopic enrichment or with a high natural isotopic abundance.

\subsection{Virtual photon excitation at RCNP}

\begin{figure}
\begin{center}
\resizebox{0.5\textwidth}{!}{%
  \includegraphics{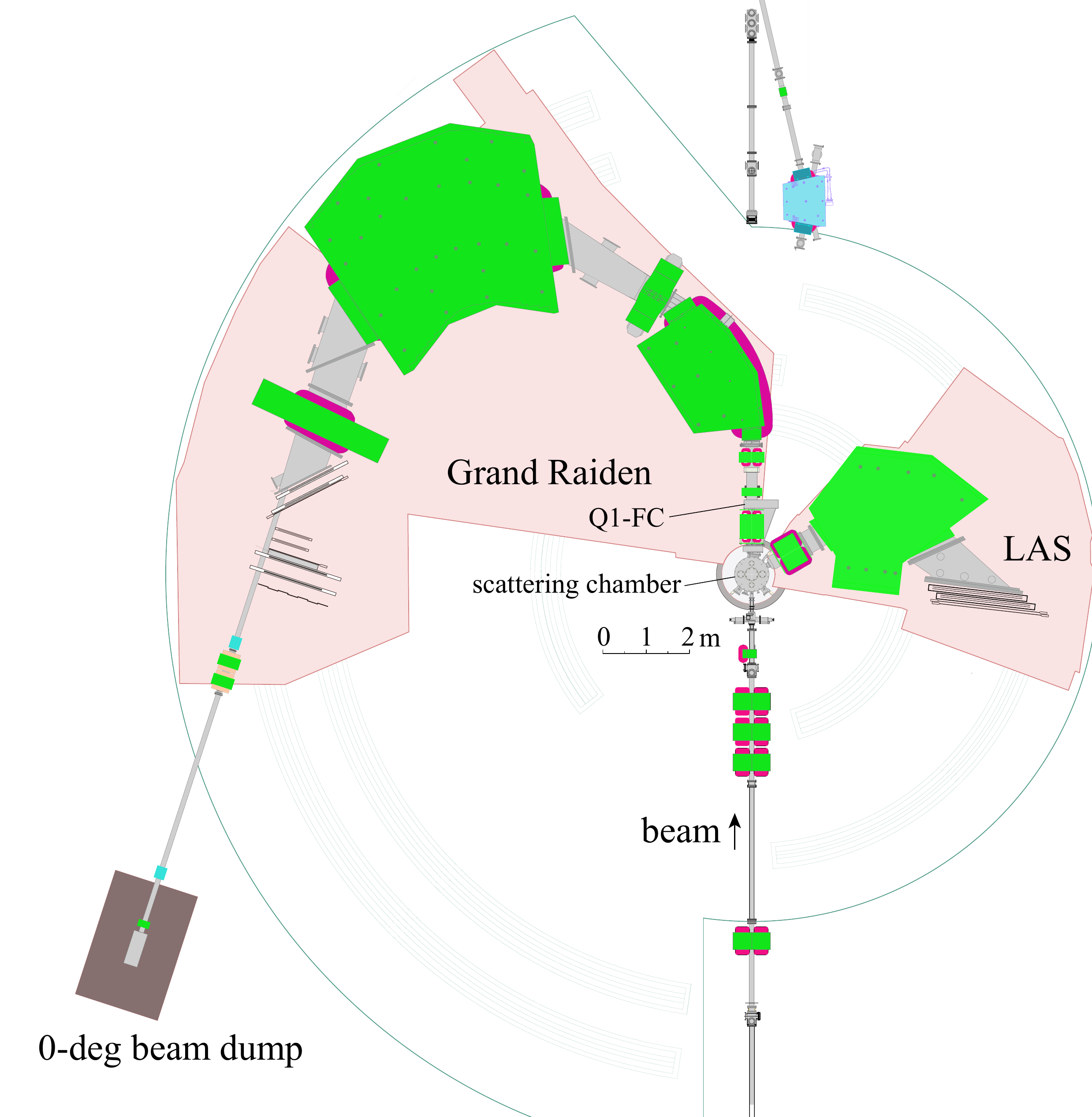}
  \includegraphics{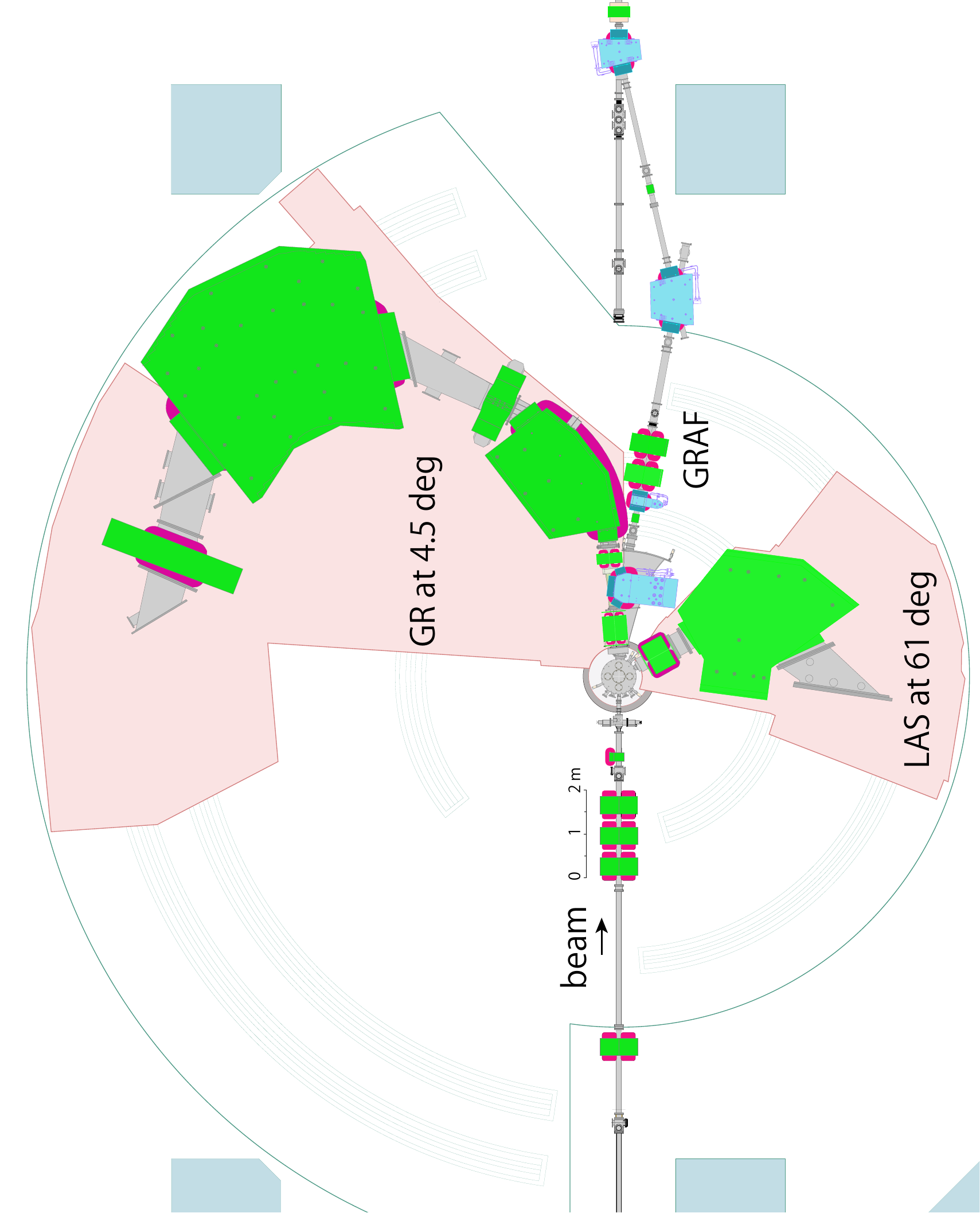}
}
\caption{
Grand Raiden spectrometer in the zero-degree inelastic scattering mode (left)~\cite{Neumann-Cosel2019electric} and in the \ac{GRAF} mode (right)~\cite{kobayashi2019}.
}
\label{fig:GRinZeroDegSetup}
\end{center}
\end{figure}

The excitation-energy distribution of the $E1$ reduced transition strength, $d B(E1)/dE_x$, will be measured using the virtual-photon excitation method at the \ac{RCNP}, Osaka University, where two cyclotrons will be used in a cascade to accelerate a proton beam to 392~MeV. 
The Grand Raiden spectrometer~\cite{fujiwara99} is placed in the zero-degree inelastic scattering mode~\cite{neu19,tam09} to measure the scattered protons at 0-3 degrees (left panel in Fig.~\ref{fig:GRinZeroDegSetup}), and at 4.5 and 6.5 degrees covering the scattering angular range of 3.5-7.5 degrees in the \ac{GRAF} mode~\cite{kobayashi2019} (right panel in Fig.~\ref{fig:GRinZeroDegSetup}).
The beam is transported to the zero-degree beam dump (wall beam dump) in the zero-degree (\ac{GRAF}) mode.
The beam dumps are well separated from the target position and are shielded by concrete to reduce radiations entering the decay detectors.
A proton beam at 392 MeV is employed to cover the excitation-energy range of 7-32 MeV in the zero-degree mode.
The details of the experimental technique can be found in Ref.~\cite{tam09}.

The multipole decomposition analysis of the $(p,p')$ reaction is well established at RCNP~\cite{Neumann-Cosel2019electric,TamiiPoltoratskaNeumannEtAl2011}.
The angular distribution of the $(p,p')$ cross sections will be fitted by a sum of theoretical angular distributions for several multipoles to extract the $E1$ component that dominates at zero degrees. The $E1$ cross section at zero degrees is converted to the E1 reduced transition strength $B(E1)$ by using Coulomb excitation calculation with the Eikonal approach.

\begin{figure}
\begin{center}
\resizebox{0.3\textwidth}{!}{%
  \includegraphics{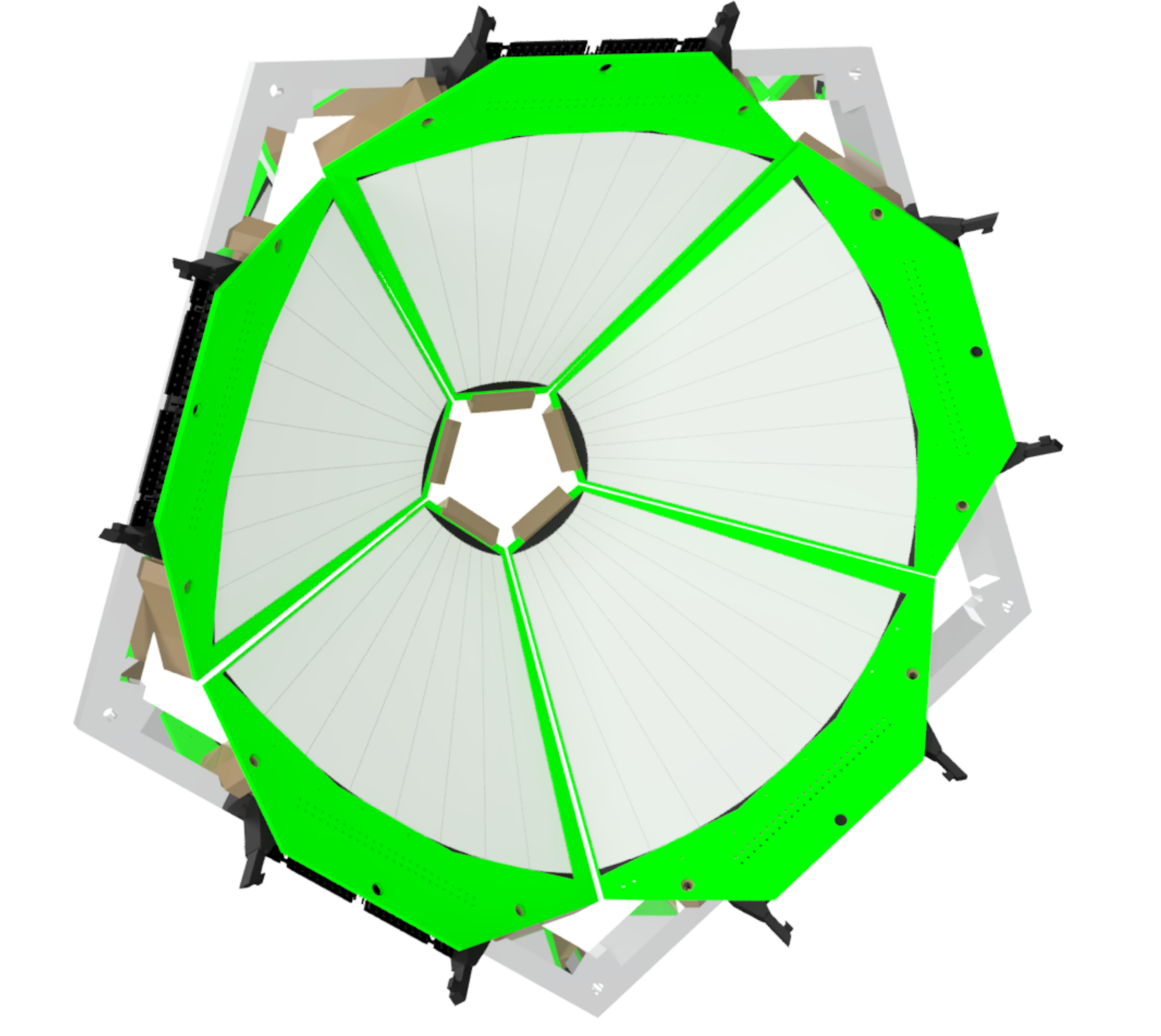}
}
\caption{
An illustrative image of a silicon detector array, \ac{SAKRA}, developed at Osaka University for charged-particle decay detection in a configuration of five~\ac{DSSSD} with a thickness of 500~$\mu$m.
}
\label{fig:SAKRA}
\end{center}
\end{figure}

A silicon detector array, \ac{SAKRA} (Si Array developed by Kyoto and osaka for Research into Alpha cluster states) \cite{sakra}, will be placed at around the target position for the coincidence measurement of decay charged particles.
\ac{SAKRA} consists of \ac{DSSSD} with a thickness of 500~$\mu$m of the MMM design from Micron Semiconductor Limited.
\ac{SAKRA} covers a solid angle of $\sim25$\% of $4\pi$ at the backward angles with respect to the beam direction. 
The lowest detectable energy is $\sim$0.5 MeV. The maximum detectable energy without penetration of the detector is 8.2, 11.0 and 33.0 MeV for $p$, $d$ and $\alpha$ particles, respectively.

\begin{figure}
\begin{center}
\resizebox{0.4\textwidth}{!}{%
  \includegraphics{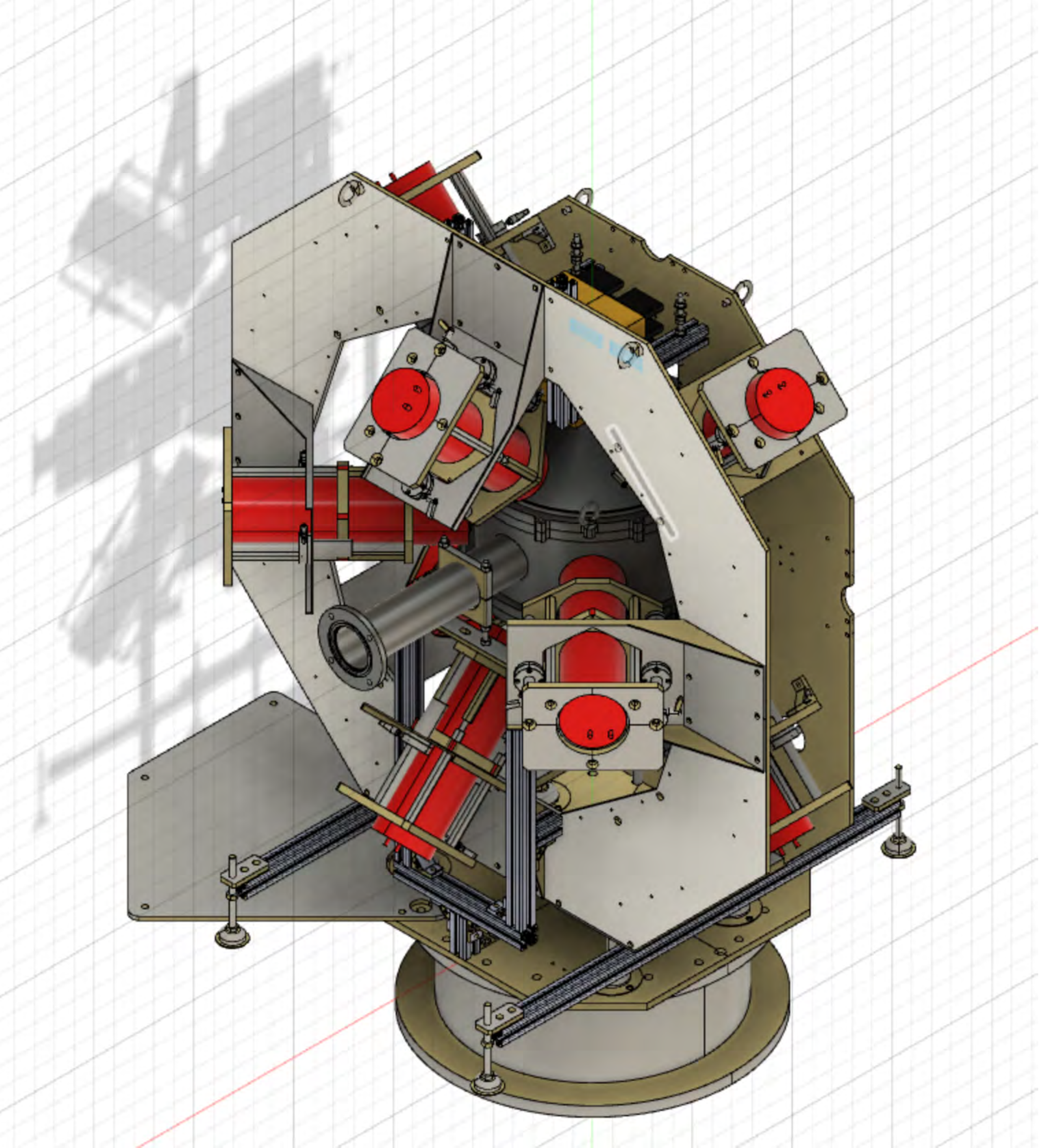}
}
\caption{
Eight large volume LaBr$_3$ detectors are placed around the target chamber that accommodates  \ac{SAKRA} in a view from the upstream side of the beam.
}
\label{fig:scglla}
\end{center}
\end{figure}

Eight large volume LaBr$_3$ detectors (3.5"$\phi$-8" L)~\cite{giaz2013,gosta2018} will be placed around the target chamber as shown in Fig.~\ref{fig:scglla}) for decay $\gamma$-ray detection~\cite{kobayashi2019}.
The distance from the target position to the front surface of the LaBr$_3$ detectors is $\sim$200 mm.
A beam-time proposal for the first PANDORA experiment at RCNP is approved and is scheduled for December 2022. The planned targets are $^{12,13}{\rm C}$, $^{27}{\rm Al}$, and $^{24,26}{\rm Mg}$ for the decay measurement in coincidence to the scattered protons and $^{10,11}{\rm B}$ only for the scattered protons.

\subsection{Virtual photon excitation at  iThemba LABS}



The dispersion-matched 200-MeV proton beam produced by the Separated Sector Cyclotron (\ac{SSC}) at iThemba LABS will be used to measure the photo-absorption response of the target nuclei.
The scattered protons will be momentum analyzed with the $K$=600 magnetic spectrometer~\cite{nev11}.
This spectrometer consists of five active elements: a quadrupole, two dipoles, and two trim coils (K and H).
The focal plane detector consists of two \ac{MWDC} measuring the horizontal and vertical position of the scattered particles, as well as two plastic scintillating detectors that provide information on the energy deposited at the focal plane.
The excitation energy can be determined from the horizontal position in the first \ac{MWDC}. 
Particle identification and halo rejection are accomplished using the energy deposited in the scintillators and the time between the trigger event and the following reference radio-frequency signal from the accelerator system.

For the \ac{PANDORA} project, the K600 spectrometer will be operated in the zero-degree and small-angle configurations in which the center of the acceptance of the spectrometer will be at 0 and 4 degrees, respectively, resulting in data acquired over a total angular range of 0$^\circ$-6$^\circ$. This will allow for the extraction of the photo-absorption cross-section from the $(p,p')$ data by applying \ac{MDA} of the measured data~\cite{Neumann-Cosel2019electric}
The \ac{MDA} is based on model predictions of the angular distribution shapes of the different multipolarities contributing to the spectra.
For each discrete transition or excitation energy bin, the experimental angular distribution is fitted employing the least-square method to a sum of the calculated angular distributions, weighted with coefficients corresponding to the different multipolarities of the transitions. Due to the increasing complexity of contributions from different multipoles at higher momentum transfers,
the \ac{MDA} is more reliable and effective in the angular range below 10$^\circ$.
Combining the zero-degree and the 4-degree configurations, 5 to 7 data points can be extracted in the 0$^\circ$-6$^\circ$ angular range.

\begin{figure}
\begin{center}
\resizebox{0.3\textwidth}{!}{%
  \includegraphics{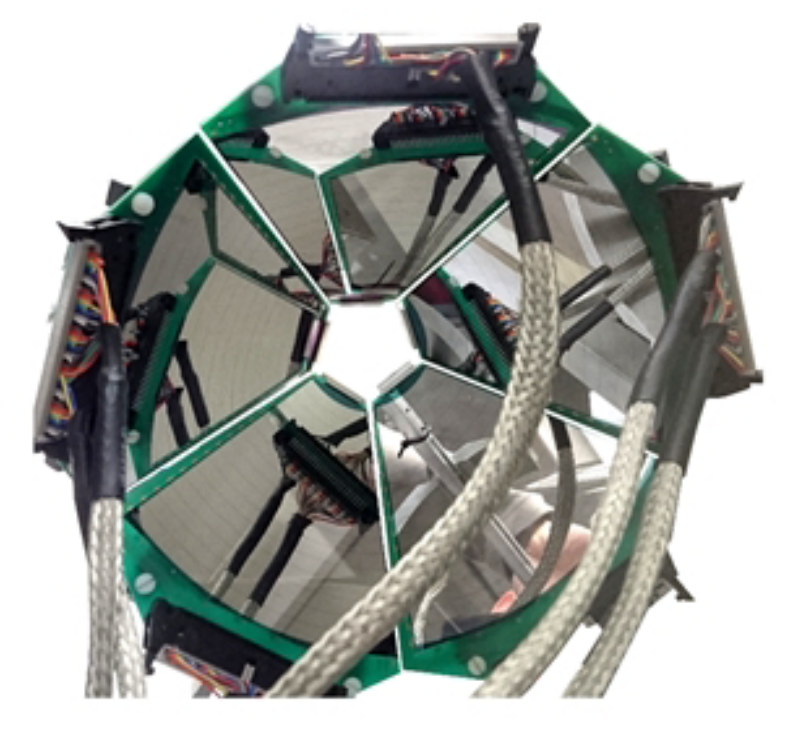}
}
\caption{
The silicon-strip detector array, CAKE, at iThemba LABS in the standard single-layer configuration.
}
\label{fig:CAKE}
\end{center}
\end{figure}

\begin{figure}
\begin{center}
\resizebox{0.5\textwidth}{!}{%
  \includegraphics{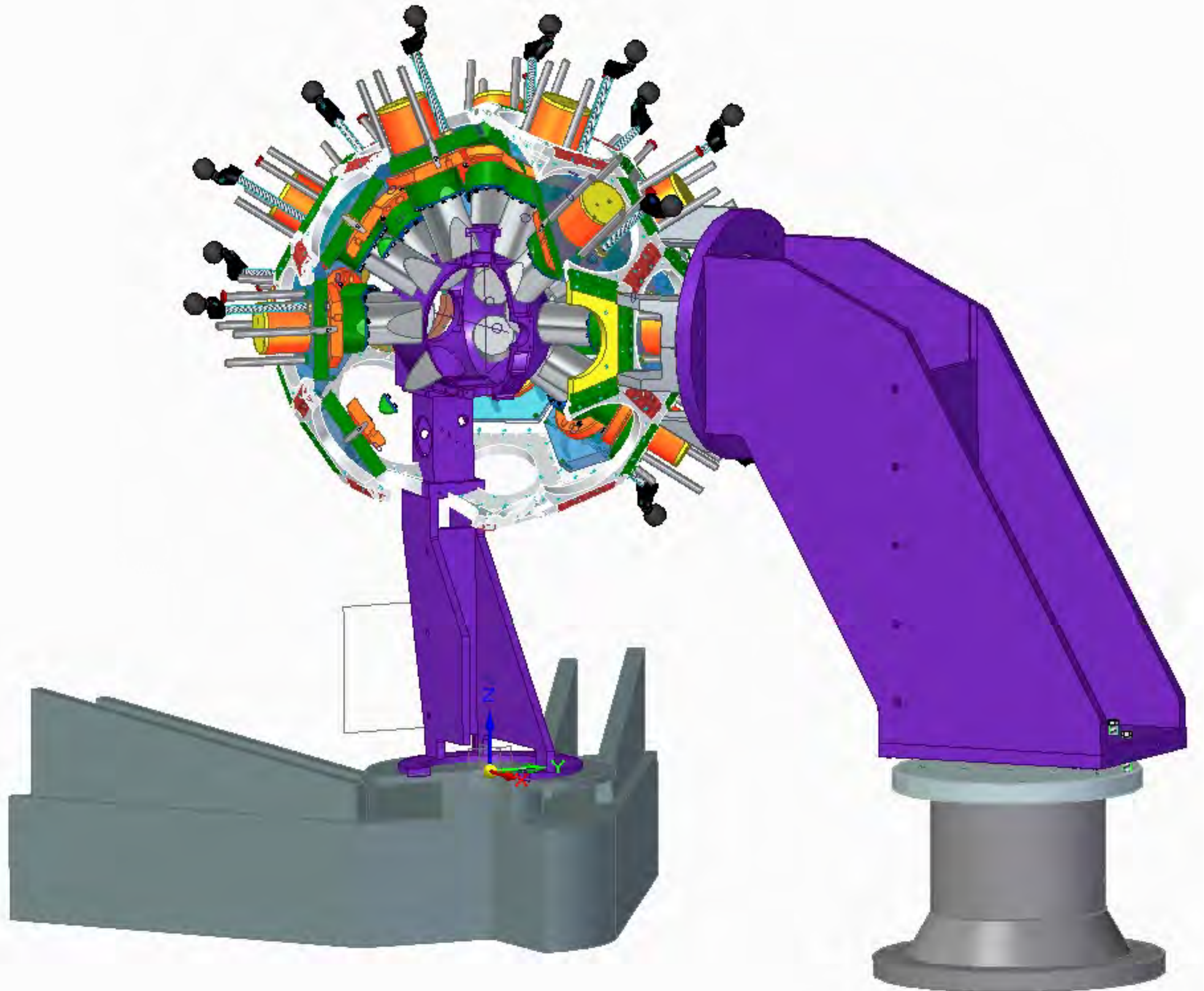}
}
\caption{Illustration of the African LaBr Array (ALBA) assembled around the scattering chamber.
}
\label{fig:ALBA}
\end{center}
\end{figure}

Coupled with the K600 spectrometer, the \ac{CAKE}~\cite{CAKE} will be used to measure the subsequent particle decays from the excitation of the \ac{GDR}.
In the standard configuration, the \ac{CAKE} comprises five \ac{DSSSD} of the MMM design from Micron Semiconductor Limited, arranged in a lamp-shaped configuration as shown in Fig.~\ref{fig:CAKE}.
The detectors are 400-$\mu$m thick and have 16 ring channels on the junction side of the detector and eight sector channels on the ohmic side.
The limitation in the proton energy detectable (7 MeV) due to the thickness of the detectors can be overcome by arranging the detectors in the double-layer configuration.
The array will be arranged for the first experiment to cover the maximum solid angle (25\%) with a single-layer configuration. One section of the array will be set up as a double layer. This will allow testing the double-layer configuration for future \ac{PANDORA} experiments.
The particle identification will be carried out using the time-of-flight technique, as shown in Ref.~\cite{CAKE}, and energy excitation energy vs silicon energy matrices can be constructed for different decay channels. The array will be coupled to the spectrometer in both angular configurations.
The measurement of all decay channels is crucial for simulating the energy and composition evolution in the propagation of the \ac{UHECR}s. Therefore the de-excitation of the \ac{GDR} by $\gamma$-ray emission will also be measured by placing large volume LaBr$_3$:Ce detectors from the African LaBr Array - \ac{ALBA} - around the scattering chamber, see Fig.~\ref{fig:ALBA}. The ALBA comprises up to 21 large volume LaBr$_3$:Ce detectors that can be arranged at an average distance of 18-20 cm from the target. These detectors will be utilized only in the K600 zero-degree configuration due to the limitation imposed by the background coming from the beam dump in the finite-angle setting.
The $\gamma$-decay information will also be used to extract the branching ratio of particle decays into excited states by detecting the $\gamma$ rays from the daughter states. 

\subsection{Real photon excitation at ELI-NP}
The focus of the \ac{ELI-NP} facility~\cite{Gales2016,Gales2018,Tanaka2020} will be to directly measure the decay strength into different channels following a population of high-energy states using real photons from a $\gamma$-ray beam.
The \ac{ELI-NP} facility has two main approaches to photo-nuclear physics.
One of these systems is the 10~PW high-power laser beam-lines~\cite{Lureau2020}.
However, for this project, the main beam-line of interest will be the high-brilliance, low-bandwidth $\gamma$-ray beam produced by Compton back-scattering of a regular laser off an electron beam.
This system, called \ac{VEGA}, is being constructed by Lyncean Technologies Inc. and will consist in an electron storage ring coupled to a Fabry-P\'{e}rot optical cavity.
The energy of the circulating electrons can be steplessly varied in the range of 234~MeV to 742~MeV and photons from two different laser configurations, covering two different energy ranges,  one at $\sim 1 $~$\mu$m and the other at $\sim 0.5 $~$\mu$m, respectively, will be able to cover $\gamma$-ray energies from 1~MeV to 19.5~MeV. 
Due to the nature of the Compton back-scattering process, the beam will be almost completely polarized. The bandwidth of the beam is predicted to be less than 0.5\%, making it an ideal probe for the type of resonant structures expected in the light nuclei studied in the \ac{PANDORA} project.
The main parameters of the \ac{VEGA} system are listed in Table~\ref{tab:vega}. 

\begin{table}[ht]
\caption{Main predicted parameters of the VEGA system at ELI-NP.
\label{tab:vega}}
 \begin{tabular*}{\columnwidth}{@{\extracolsep{\fill}}lc}
\hline
Parameter & Limit\\
\hline
Energy & $\leq 19.5$ MeV \\
Polarisation & $\geq95$\%  \\
Bandwidth (FWHM) & $\leq 0.5$\%  \\
Peak Spectral Density & $\geq 5000$ s$^{-1}$eV$^{-1}$  \\
Off-Peak Background Density & $\leq 1.0 \times 10^{-2}$ s$^{-1}$eV$^{-1}$\\
Divergence (FWHM) & $\leq 1.5\times10^{-4}$ rad  \\
Repetition rate & $\lesssim 72$ MHz  \\
Beam intensity at 10~MeV & $\sim 2.5\times10^{8}$ s$^{-1}$  \\
\hline
 \end{tabular*} 
\end{table}

\begin{figure}
\begin{center}
\resizebox{0.5\textwidth}{!}{%
  \includegraphics{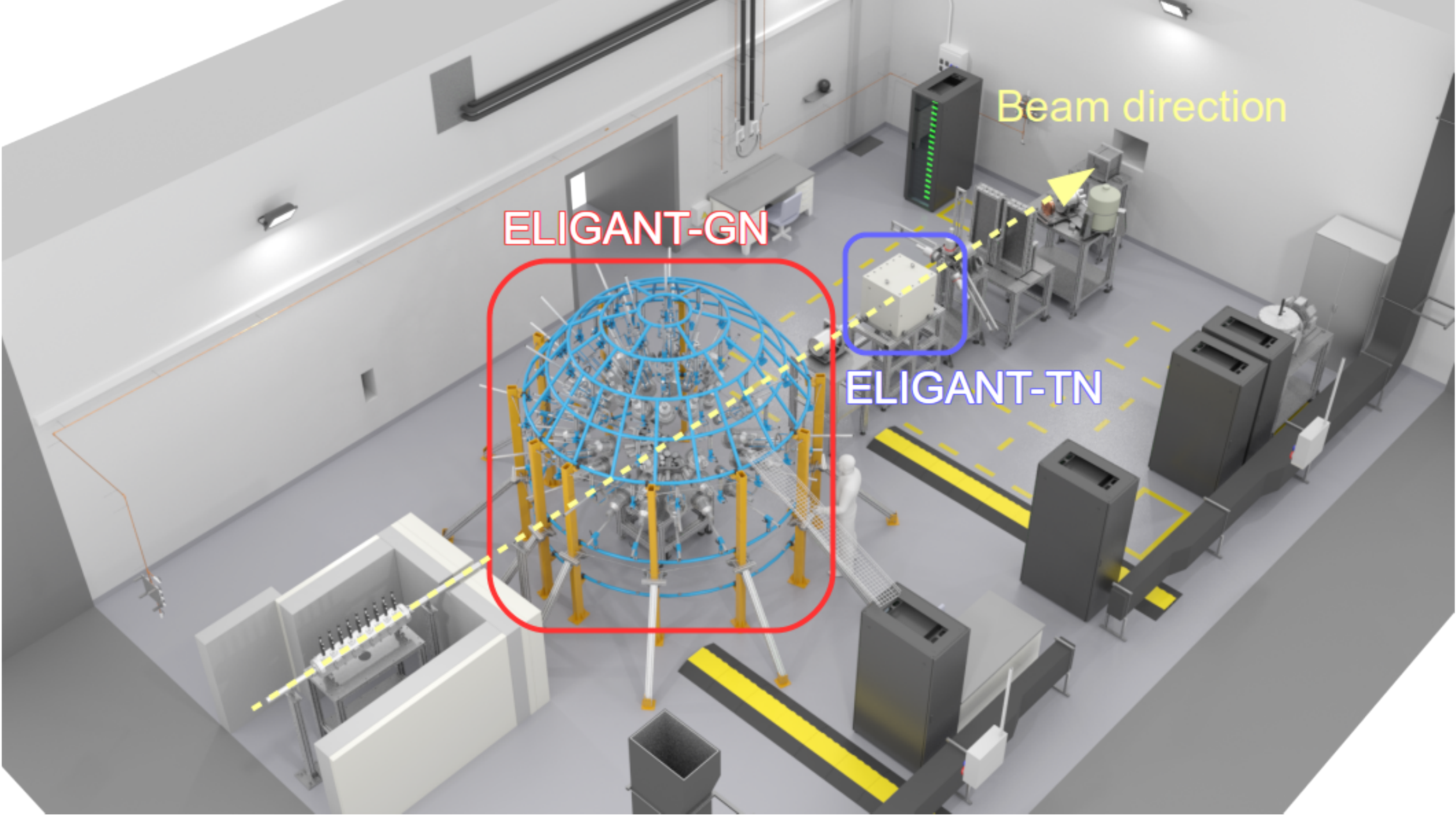}
}
\caption{ELI-NP experimental area for experiments above neutron threshold.}
\label{fig:e9}
\end{center}
\end{figure}

Several different experimental setups are currently being prepared for photo-nuclear measurements at \ac{ELI-NP} related to nuclear structure and photo-nuclear reactions.
The reaction studies, with the most significant relevance to the project discussed here, consist of both charged particle and neutron-detection following the nuclear excitations with the $\gamma$-ray beam.
For neutron detection, two different setups have been constructed at the \ac{ELI-NP} facilities: the \ac{ELIGANT-TN} setup \cite{Camera2016} for high-precision cross-section measurements for $(\gamma,x\mathrm{n})$ reactions, and \ac{ELIGANT-GN} \cite{Camera2016,Krzysiek2019a,Soderstrom2022} aiming for detailed investigation of correlations and competition between the neutron and $\gamma$-ray decay channels from states above the neutron threshold.
The experimental area is shown in Fig.~\ref{fig:e9}.

\ac{ELIGANT-GN} consists of 34 large-volume LaBr$_{3}$:Ce and CeBr$_{3}$ detectors with a 76~mm length and 76~mm diameter for high-energy $\gamma$ rays.
For neutron detection, 36 liquid scintillator detectors of EJ-301 type at a distance of 1.5~m from the target will be used for high-energy neutrons, and 25 lithium-glass detectors of type GS20 at a distance of 1~m from the target will be used for low-energy neutrons.
The role of the large-volume crystal scintillators is twofold.
One of their purposes is to identify the weak one-step and two-step branchings from the \ac{GDR} states. The other one is to give a clear time reference for neutron-energy measurements via time-of-flight in the liquid- and glass scintillators.
In a recent commissioning paper \cite{Soderstrom2022}, the total $\gamma$-ray detection efficiency is estimated to be around 1\% at an energy of 10~MeV for the $\gamma$ rays.
The liquid scintillators have an estimated efficiency of around 1\% at a neutron energy of 10~MeV, and up to 2.5\% for a neutron energy of 2~MeV.
Below 1~MeV, the efficiency drops rapidly. However, the inclusion of the lithium-glass scintillators can provide low-energy neutron detection with an efficiency of about 0.5\% at a neutron energy of 250~keV.

\ac{ELIGANT-TN} consists of 28 tubes filled with $^{3}{\rm He}$, embedded in a polyethylene matrix.
The efficiency of the neutron counter is around 38\% over the complete, predicted, neutron energy range. 
This feature is essential as it means the energy spectrum of the emitted neutrons does not bias the measured cross-section.
However, although the detector is used with neutrons moderated to thermal energies, the average neutron energy can still be obtained from the ring-ratio method~\cite{Berman1975}, giving additional information for comparison with different theoretical approaches.
Due to the flat efficiency, higher-order neutron emission cross-sections can be disentangled from the total neutron-emission cross-section using the direct multiplicity sorting procedure, which is well described in References~\cite{Gheorghe2017,Utsunomiya2017,Gheorghe2021}.

A common assumption in the measurements of the \ac{GDR} cross-section for heavy nuclei is that the total decay of the \ac{GDR} is very close to the neutron decay channel. 
This, however, is not necessarily the case for the nuclei to be studied in the \ac{PANDORA} project.
Besides internal transition decay via $\gamma$ rays, a significant part of the decay could go via charged particles like protons or $\alpha$ particles. At \ac{ELI-NP}, the charged particle reaction studies will primarily be carried out with the silicon-strip detector array, \ac{ELISSA}, and the large time-projection chamber ELI-TPC \cite{Tesileanu2016}.
For experiments within the \ac{PANDORA} project, the ELI-TPC can be operated in stand-alone mode for rare reaction channels with gas targets. \ac{ELISSA}, in the configuration proposed in the Technical Design Report, will consist of three rings of twelve $X3$ position-sensitive detectors \cite{Chesnevskaya2018} in a barrel-like configuration, and two end cap assemblies.
In order to integrate \ac{ELISSA} with \ac{ELIGANT-GN}, a new compact configuration will be required.
This, however, will allow us to also investigate multiple-particle reactions, for example $(\gamma,\mathrm{np})$ or $(\gamma,\mathrm{n}\alpha)$ in the cases of $\alpha$-cluster nuclei with excess neutrons such as $^{13}$C. 

For this project, the $\gamma$-ray beam as provided by \ac{ELI-NP} will be used with approximately 50-100~keV energy resolution in the energy range from particle separation thresholds up to the maximum available $\gamma$-beam energy. 
The targets proposed are neither very rare nor radioactive we are not limited by the beam size at the experimental point and can accept beam spots of a few millimeters.
For the same reason, we are not very sensitive to the beam intensity and can accept a lower intensity by introducing a time structure in the beam. This sacrifice in intensity can easily be countered by using thicker targets and would have the additional benefit of increased accuracy and precision in the neutron energy measurements, as well as adding the possibility to separate $(\gamma,1\mathrm{n})$ and $(\gamma,2\mathrm{n})$ cross sections in \ac{ELIGANT-TN} for events above the two-neutron separation threshold.
The use of fully polarized beams allows to separate the $E1$ response and the $M1$ response, as well as to separate $s$-wave and $p$-wave neutron emission.
For a given isotope, different thicknesses of the targets can be produced for different energies keeping the total reaction rate as constant as possible in the detector setup.
This will ensure that the statistical uncertainties remain similar and that systematic uncertainties related to the count rate in the detector system are minimized. 

\subsection{Other related facilities}

\subsubsection{HI$\gamma$S}

\begin{figure}
\begin{center}
\resizebox{0.5\textwidth}{!}{%
  \includegraphics{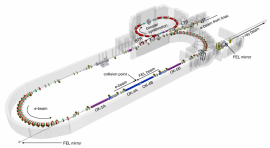}
}
\caption{
Scheme of the HI$\gamma$S facility. For details, see Ref.~\cite{Weller09}. Reprinted with permission from~\cite{Weller09}.
}
\label{fig:higs_setup}
\end{center}
\end{figure}

The High Intensity $\gamma$-ray Source (HI$\gamma$S)~\cite{Weller09} is a joint
project between Triangle Universities Nuclear Laboratory (TUNL) and the Duke
Free Electron Laser Laboratory (DFELL). The HI$\gamma$S facility is shown
schematically in Fig. \ref{fig:higs_setup}. An electron beam is generated in a
photo-cathode microwave electron gun, bunched, and pre-accelerated to energies of
$E_{e^{-}}$ = 0.18 to 0.28 GeV in an electron linear accelerator. The subsequent booster
synchrotron can ramp up the energy up to 1.2 GeV before the electron
bunches are injected into the Duke electron storage ring. Within the storage
ring a free electron laser (FEL) is powered by the electron beam.

The ring electrons are deflected by several wiggler magnets
placed in the storage ring and emit horizontally or circularly polarized FEL photons
with variable wavelengths from 190~nm to 1060~nm. These photons are reflected by
the FEL mirrors and collide with another electron bunch at the collision
point. In the laser Compton backscattering (LCB) process, the photon energy
can be boosted up to 100 MeV with a total flux on target in the order of $\sim
10^{8}$ $\gamma/s$ depending on the scattering angle. The LCB beam energy can
be tuned by adjusting the electron and FEL energy, respectively.
Due to the polarization conservation of the Compton backscattering process, a
linearly or circularly polarized high-energy LCB photon beam is produced. The
high-intensity photon beam passes the FEL mirror and is collimated about 60~m
downstream of the collision point, respectively. Depending on the spatial distribution of the
backscattered photons and the collimator size, the typical full width at half
maximum (FWHM) of the beam energy profile is around 1~\% to 3~\% of the peak
energy.

Two experimental halls with dedicated detector setups~\cite{Loeher13} are located in the Upstream Target Room (UTR) and the Gamma Vault, which use the high-energy photon beam to perform nuclear physics experiments. A number of NRF studies have been performed
in the past decades to determine spin- and parity quantum numbers of excited
states~\cite{Pietralla02,Isaak11,Goddard13,Schwengner21}, investigate
collective excitation modes in deformed nuclei \cite{Beck17,Beck20,Ide21,Kleemann21}, extract photoabsorption cross sections~\cite{Tonchev10,Loeher16,Isaak13,Isaak21} and photon strength functions~\cite{Isaak19,Papst20}. 
Furthermore, the measurement of ($\gamma$, n) reactions in light nuclei~\cite{Tornow03,Ahmed08,Zimmerman13} and in the Fe-Ni region were performed~\cite{Scheck13,Schwengner20}. 
Very recently, the $\gamma$-decay of the IVGDR into the ground-state rotational
band in the deformed nucleus $^{154}$Sm was observed~\cite{KleemannPC} and demonstrated the
the potential of LCB beams for the investigation of photoabsorption cross-sections
and decay channels for the physics goals in PANDORA.
For a recent review article on nuclear structure studies and
applications with a photonuclear reaction below and above particle emission
thresholds see Ref.~\cite{Zilges22}.

\subsubsection{NEPTUN at S-DALINAC}

\begin{figure}
\begin{center}
\resizebox{0.4\textwidth}{!}{%
  \includegraphics{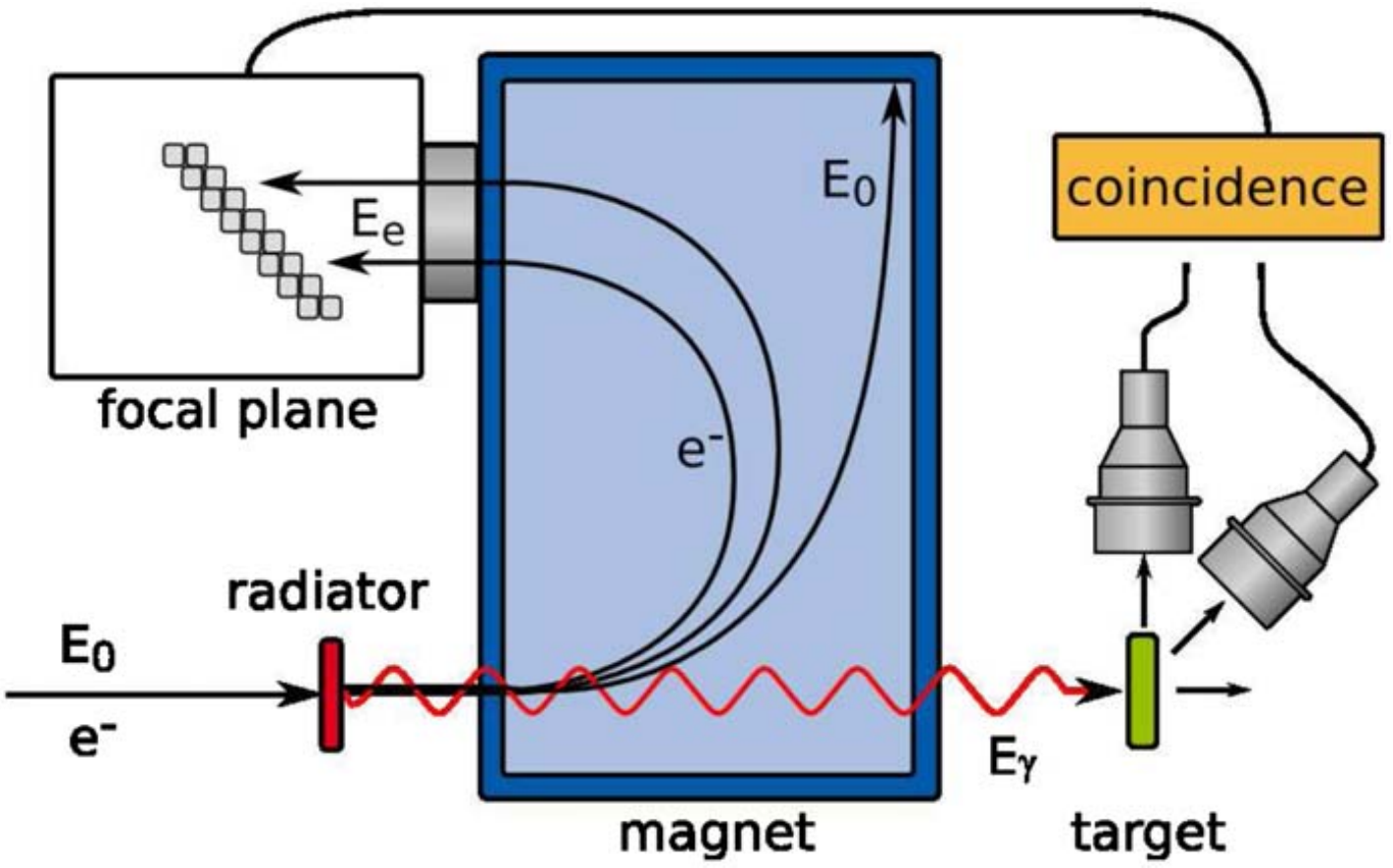}
}
\caption{Principle of photon tagging. For details, see Ref. \cite{Savran10}. Reprinted figure with permission from \cite{Savran10}.
}
\label{fig:neptun_setup}
\end{center}
\end{figure}

The photon-tagger NEPTUN \cite{Savran10} operated at the S-DALINAC \cite{Pietralla18}
of the Institute for Nuclear Physics of TU Darmstadt provides a beam of
energy-tagged and, therefore, quasi-monoenergetic photons. The basic principle is
illustrated in Fig.~\ref{fig:neptun_setup}. 

The electron beam of
the S-DALINAC with energy $E_0$ impinges a thin radiator target producing bremsstrahlung photons.
The reacted electrons are momentum-analyzed via dispersion in a large dipole
magnet and detected by a focal plane consisting of scintillator strips coupled
to silicon photomultiplier. The energy of the generated photon ($E_{\gamma} =
E_{0} - E_{e}$) is determined from the remaining energy of the reacted electron
$E_e$ and $E_0$ neglecting the recoil energy transferred to the atom
in the radiator in the bremsstrahlung process.
The coincident measurement of the induced photonuclear reactions on the samples
in the experimental setup, and the reacted electron enables to 'tag' the initial
photon energy and thus conduct experiments with quasi-monoenergetic photon
beams. Tagged photons in the energy range from 5 to 35 MeV can be produced with maximum
spectral photon densities of 103 photons/keV/s after collimation with a solid angle of about 3 mrad. 
The experimental setups are located about 5 m downstream of the photon tagger
NEPTUN. Currently, two possible setups exist, namely the GALATEA LaBr-detector
array to detect high-energy photons and the fast target
changer PROTEUS. The latter is combined with a large-volume CeBr
scintillator positioned in-beam downstream of PROTEUS to directly measure photoabsorption cross sections via total photoabsorption. The advantage is
that the complete absorption is measured directly and independently of
any particular exit channel. Thus, future experiments at NEPTUN and PROTEUS have
the potential to provide complementary datasets for the photoabsorption cross sections in the
mass region studied in PANDORA.

\subsubsection{SLEGS at SSRF}

\begin{figure}
\begin{center}
\resizebox{0.5\textwidth}{!}{%
  \includegraphics{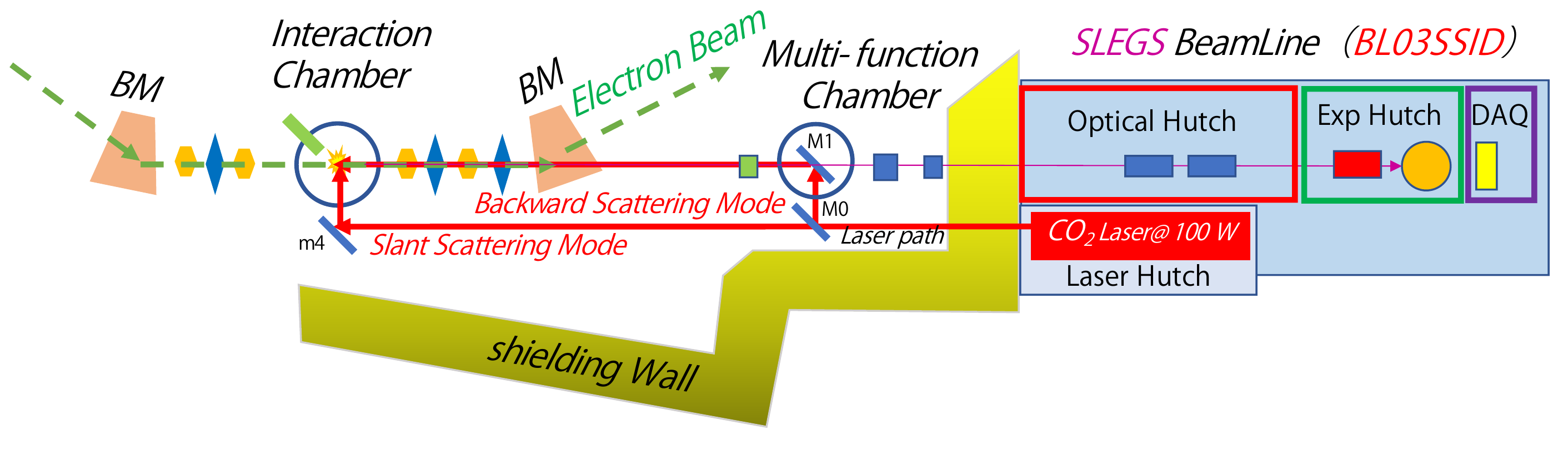}
}
\caption{
The $\gamma$-ray production beam-line, SLEGS, built at the Shanghai Synchrotron Radiation Facility (SSRF). 
}
\label{fig:SLEGS}
\end{center}
\end{figure}

The \ac{SLEGS} depicted in Fig.~\ref{fig:SLEGS} is the $\gamma$-ray production beam line built at the Shanghai Synchrotron Radiation Facility (SSRF). The slant-scattering mode is for the first time employed to systematically produce energy-tunable $\gamma$-ray beams in \ac{LCS} of 10.64$\mu$m photons from a CO$_2$ laser with 3.5 GeV electrons in the storage ring. The \ac{SLEGS} officially completed its commissioning run between July and December 2021 \cite{Wang2022}. The interaction chamber dedicated to the slant-scattering was constructed \cite{Xu2022}. The energy range of \ac{LCS} $\gamma$-rays produced at \ac{SLEGS} is 0.66 - 21.7 MeV. The maximum energy of the $\gamma$-ray beam is 0.66 MeV at the slant-scattering angle 20$^{\circ}$ and 21.1 MeV at 160$^{\circ}$, while the highest achievable energy is 21.7 MeV in the back-scattering at 180$^{\circ}$. The CO$_2$ laser is vertically introduced into the interaction chamber, transported through a set of mirrors and convex lenses, and focused at the collision point with 3.5 GeV electrons. The laser optical elements are mounted on a turn table inside the interaction chamber to continuously change the slant-scattering angle from 160$^{\circ}$ to 20$^{\circ}$. The $\gamma$-ray flux at the production point is 2.1 $\times$ 10$^4$ at 20$^{\circ}$, 2.5 $\times$ 10$^6$ at 90$^{\circ}$, and 1.2 $\times$ 10$^7$ at 180$^{\circ}$ for the 5 W (maximum 100 W) CO$_2$ laser and 300 A electron beams.  $\gamma$-rays are led to the experimental hutch through the double collimator system \cite{Hao2021}. The energy resolution of $\gamma$-ray beams at the target position is 2 - 15 \% in the full width at half maximum, depending on the slant-scattering angle, provided that the $\gamma$-ray emission angle is confined to be less than 0.45 mr with the collimator system. Four types of detectors meet the versatility requirements of nuclear physics experiments. This makes it possible to perform nuclear resonance fluorescence, flat-efficiency neutron, high-resolution neutron time-of-flight, and low-energy charged-particle measurements for photonuclear reactions \cite{Wang2022}. In the \ac{PANDORA} project, from the iron group (Fe-Co-Ni) to lithium nuclei, one can investigate the neutron decay of the giant- and pygmy-dipole resonances, mainly $(\gamma,1n)$ cross sections, decay branching ratios to excited states in the residual nuclei, and the charged-particle decay, mainly $(\gamma, p)$ and $(\gamma,\alpha)$ cross sections.

\subsubsection{The SCRIT electron scattering facility for short-lived exotic nuclei}

\begin{figure}[hb]
\begin{center}
\resizebox{0.4\textwidth}{!}{%
  \includegraphics{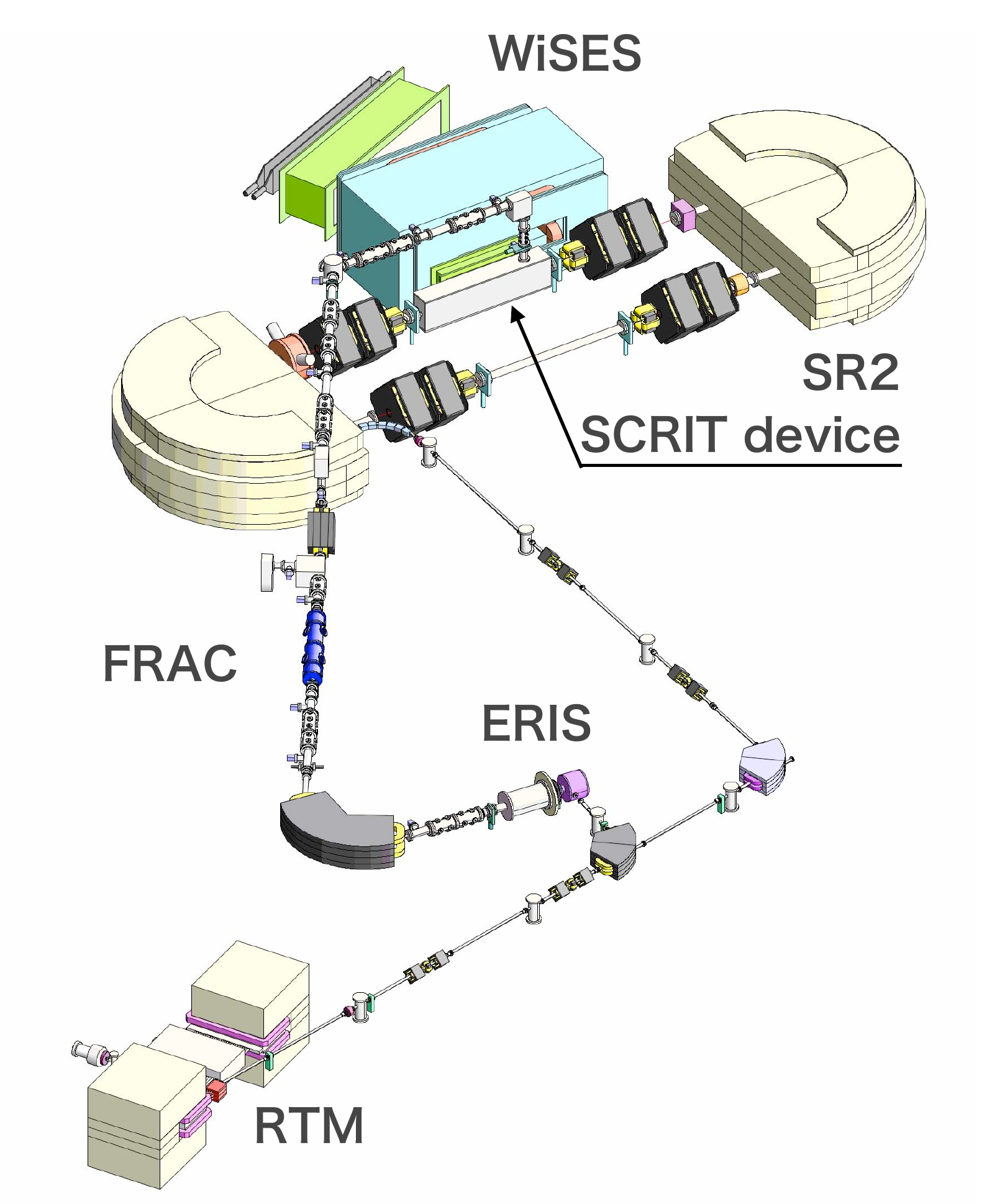}
}
\caption{SCRIT electron scattering facility at the RIKEN RI Beam Factory.}
\label{fig:SCRITfacility}
\end{center}
\end{figure}

A ground-breaking electron scattering facility has been built at RIKEN, Japan, and is currently in operation. The \ac{SCRIT} electron scattering facility is dedicated to short-lived exotic nuclei far from stability, shown in Figure \ref{fig:SCRITfacility}.

Structure studies of exotic nuclei by electron scattering are possible at the \ac{SCRIT} facility thanks to the novel target forming technique combined with a high-energy electron beam~\cite{Wakasugi2004}. The \ac{SCRIT} technique provides high luminosity, $\sim 10^{27} \mathrm{cm}^{-2} \mathrm{s}^{-1}$, enough for elastic scattering with a very small number of target ions, typically $10^9  \mathrm{cm}^{-2}$~\cite{Tsukada2017}. 

In addition to the ground-state charge density profile of short-lived exotic nuclei by elastic electron scattering, their $E1$ responses will also be accessible at the \ac{SCRIT} facility. The $E1$ responses of exotic nuclei have been studied only by Coulomb excitation reactions in nucleus-nucleus collisions. Since it is not a purely electromagnetic reaction, its interpretation has been questioned. 

The inelastic electron-scattering cross section at the ultra-forward angle, $\theta \sim 0 ^\circ$, is known to be related to the photo-nuclear cross section by virtual-photon theory. Detecting the inelastic electron scattering off exotic nuclei at a very forward angle allows for determining the total photo-absorption cross-section. It should be stressed that at this facility, one can cover the full \ac{GDR} region with an electron beam energy of $\sim 100$ MeV~\cite{Suda2017}, previously limited to a few cases even using energetic RI beams.

\section{Theoretical Nuclear Physics}
\label{sec:theory}


In the nucleus rest frame, at typical \ac{UHECR} energies of 10$^{19}$-10$^{21}$ eV, the \ac{CMB} photons are boosted to energies in a range between a few hundreds of keV up to a few hundreds of MeV. The interaction process between the \ac{UHECR}s and the \ac{CMB} is dominated by the \ac{GDR} at photon energies below 30-50~MeV, and to a lesser extent, by the quasideuteron emission for intermediate energies (between 50 MeV and 150 MeV) and the pion photoproduction at energies above 150 MeV \cite{all12}.
Nuclei are photo-disintegrated by emitting nucleons through ($\gamma$,n), ($\gamma$,p), ($\gamma$,2n), and other reactions. During the photo-disintegration path from Iron to protons, it is necessary to describe the dipole strength of the involved nuclei accurately. It should be noted that a large number of nuclei along this path are unstable.

The photoreaction cross sections can be estimated with the \talys\ nuclear reaction code \cite{kon04},
which takes into account all types of direct, pre-equilibrium, and compound mechanisms to estimate the total reaction probability and the competition between the various open channels. For this purpose, the photoreaction cross section is usually estimated at energies up to 50 MeV. Using a large variety of nuclear structure models to describe such cross sections is relevant, as it will allow us to estimate the corresponding typical theoretical error.

\subsection{RPA-EDF}


The \ac{RPA} with an effective \ac{EDF} is one of the standard tools to calculate the $B(E1)$ distributions of nuclei.
The \ac{RPA}-\ac{EDF} describes $E1$ and other modes, such as monopole or quadrupole, in nuclei covering the entire nuclear chart for each effective \ac{EDF}. 
They provide good systematic reproduction of experimental data, including the collective properties of the giant resonances.
They also predict \ac{PDR}, also called low-lying electric dipole (LED) mode, in a wide mass region.  

The \ac{RPA} equation with \ac{EDF} is derived as a small-amplitude vibration or a linear response to external perturbations of the time-dependent \ac{DFT}~\cite{RingSchuck1980}. Based on the \ac{DFT} that contains only one-body densities, the \ac{RPA}-\ac{EDF} equation is formulated from the one-body densities calculated from the single-particle wave functions of an $A$ nucleon system. Compared to other calculations, the approach dramatically reduces the computational cost and applies to heavy nuclei and infinite nuclear matter.
Its predictability is less reliable for very light nuclei, typically $A < 10$, than for heavy nuclei.

In the \ac{RPA}-\ac{EDF}, collective vibrational modes are described as a superposition of single particle-hole excitations providing a straightforward interpretation of the properties of these modes. On the other hand, due to a lack of many-particle-many-hole excitations, the \ac{RPA}-\ac{EDF} cannot describe the spreading width, failing the reproduction of the width of giant resonances. Spreading width of 1-2 MeV is often assumed to reproduce the experimental widths.

The \ac{RPA}-\ac{EDF} is powerful for describing the photo-nuclear cross-section that will be measured in the PANDORA project.
Developments to improve the predictions for light nuclei are in progress. 
A Monte Carlo calculation has recently shown that a few Skyrme-\ac{EDF} parameters are correlated with the peak energy of \ac{GDR} \cite{Inakura22}.
This would enable us to optimize the Skyrme-\ac{EDF} parameters to reproduce the peak energy of the measured \ac{GDR}. 
We plan to develop a new set of Skyrme-\ac{EDF} parameters that better describe the $E1$ strength distribution than the one presently available.

\subsection{Relativistic nuclear field theory (RNFT)}
\label{RNFT}


The \ac{RNFT} was developed throughout the last couple of decades as a response to novel challenges in nuclear structure and astrophysics. The \ac{RNFT} emerged as a synthesis of the late extensions of the Landau-Migdal Fermi liquid theory, Copenhagen-Milano \ac{NFT} and \ac{QHD} \cite{LitvinovaRing2006,LitvinovaRingTselyaev2007,LitvinovaRingTselyaev2008,LitvinovaRingTselyaev2010,Litvinova2015,RobinLitvinova2016,Robin2019}.
The \ac{QHD}, being a covariant theory of interacting nucleons and mesons, constrained by the low-energy \ac{QCD}, turned out to be very successful on the mean-field level \cite{BogutaBodmer1977,SerotWalecka1979,Serot1986a,SerotWalecka1997,Ring1996}. The idea of fine-tuning meson masses and coupling constants and introducing a non-linear scalar meson led to an excellent quantitative description of nuclear ground states. Thus, \ac{QHD} has provided the connection between the low-energy \ac{QCD} scale and the nucleonic scale of complex strongly-interacting media. The time-dependent version of the \ac{RMF} model and the response theory built on it have allowed for a very good description of the positions of collective vibrational states in the \ac{RRPA} \cite{RingMaVanGiaiEtAl2001,VretenarAfanasjevLalazissisEtAl2005,LiangVanGiaiMeng2008}  or, for the superfluid systems, by the \ac{RQRPA} \cite{PaarRingNiksicEtAl2003,PaarNiksicVretenarEtAl2004a}. The \ac{RMF} and R(Q)RPA form the content of the \ac{CDFT}, which performs very well and provides, in general, a better description of nuclear properties than non-relativistic \ac{DFT}'s. 

First, the many-body correlations associated with temporal non-localities of the nucleonic self-energy and effective interaction were taken into account in the extensions of the \ac{CDFT}. Namely, the first step connecting single-nucleon and vibrational degrees of freedom was made by the \ac{RNFT} \cite{LitvinovaRing2006,LitvinovaRingTselyaev2007,LitvinovaRingTselyaev2008,LitvinovaRingTselyaev2010,Litvinova2015,Litvinova2016}, a relativistic version of the original \ac{NFT} \cite{BesBrogliaDusselEtAl1976,BortignonBrogliaBesEtAl1977,BertschBortignonBroglia1983,MahauxBortignonBrogliaEtAl1985,ColoBortignon2001,NiuColoVigezziEtAl2014,NiuNiuColoEtAl2015} and the extended theory of finite Fermi systems \cite{KamerdzhievTertychnyiTselyaev1997,KamerdzhievSpethTertychny2004},
which accounts for in-medium retardation effects of the meson exchange, missing in \ac{CDFT}, in the leading approximation. The latter is based on the emergence of collective degrees of freedom, such as vibrations (phonons) caused by coherent nucleonic oscillations. An order parameter associated with the \ac{qPVC} vertices provides consistent power counting and controlled truncation schemes. In the implementations with effective interactions, the \ac{qPVC} vertices and frequencies, which give the most important contributions to the nucleonic self-energy and effective interaction beyond the \ac{CDFT}, can be well approximated by the R(Q)RPA. The non-perturbative treatment of the \ac{qPVC} effects is based on the time ordering of two-loop and higher-order diagrams, containing multiple exchanges of vibrations between nucleons, and evaluating their contributions to the one- and two-nucleon propagators. The leading approximation to the nucleonic self-energy includes 'one (quasi)  particle coupled to one phonon' $1q\otimes phonon$ configurations \cite{LitvinovaRing2006,Litvinova2012}, while 'two (quasi) particles coupled to one phonon' $2q\otimes phonon$ or $2phonon$ configurations enter the extended induced interaction
\cite{LitvinovaRingTselyaev2008,LitvinovaRingTselyaev2010}. Later generalizations of the \ac{RNFT} were devoted to configurations of higher complexity (np-nh, or $2q\otimes$Nphonon) \cite{Litvinova2015}, the inclusion of isospin-flip phonons \cite{Litvinova2016,RobinLitvinova2018}, ground state correlations caused by \ac{qPVC} \cite{Robin2019}, and finite-temperature effects \cite{LitvinovaWibowo2019,LitvinovaRobinWibowo2020}.

The nuclear response theory with the \ac{qPVC} effects, which was developed within this formalism in a parameter-free way and called \ac{RQTBA}, has provided a high-quality description of gross properties of the giant resonances \cite{LitvinovaRingTselyaev2008,MarketinLitvinovaVretenarEtAl2012,LitvinovaBrownFangEtAl2014} and some fine features of excitation spectra at low energies \cite{EndresLitvinovaSavranEtAl2010,RobinLitvinova2016} in both neutral and charge-exchange channels for medium-mass and heavy nuclei.  In particular, the isospin splitting of the pygmy dipole resonance has been explained quantitatively \cite{EndresLitvinovaSavranEtAl2010} and the beta-decay half-lives were reproduced very successfully already in the first version of the proton-neutron \ac{RQTBA} \cite{RobinLitvinova2016}. The generalized \ac{RQTBA} with multiphonon couplings \cite{Litvinova2015} opened a way to unify the theory of high-frequency collective oscillations and low-energy spectroscopy \cite{LitvinovaSchuck2019}.
The essential features of the \ac{RNFT} are that (i) it is constrained by the fundamental underlying theory, such as \ac{QCD}, and hence, consistent with Lorentz invariance, parity invariance, electromagnetic gauge invariance, isospin and chiral symmetry (spontaneously broken) of \ac{QCD};
(ii) it includes effects of nuclear superfluidity on equal footing with the meson exchange and \ac{qPVC}, so that it applies to open-shell nuclei; (iii) being a parameter-free extension of the \ac{CDFT}, it is applicable and demonstrates a high quality of performance throughout almost the entire nuclear chart, from the oxygen mass region to superheavy nuclei \cite{LitvinovaAfanasjev2011,Litvinova2012}. 

Lately, the R(Q)TBA was re-derived in the model-independent \ac{EOM} framework based on the bare fermionic Hamiltonian without applying time-blocking operators \cite{LitvinovaSchuck2019}. The \ac{EOM} formalism allowed for an {\em ab initio} description and further extensions of the \ac{RNFT}. Overall, the original and extended versions of the R(Q)TBA demonstrated significant improvements in the description of nuclear collective excitations and soft modes compared to the standard (Q)RPA theories. 

\subsection{EDF-QPM}


A microscopic approach based on self-consistent \ac{EDF} and \ac{QRPA} formalism complemented by multiphonon configurations has been developed and used in various studies of the nuclear structure of low-energy excited states, pygmy and giant resonances \cite{Tso04,Tso08}. Currently, the \ac{EDF} + three-phonon \ac{QPM} theory \cite{Sol76,Tso16} is a successful method that allows for a unified description of low-energy single-particle and multiphonon states and giant resonances. For example, a uniform treatment is required to separate between multiphonon states, \ac{PDR} and \ac{GDR} strengths. An essential advantage of the \ac{QPM} compared to other methods with quasiparticle-phonon coupling is the use of sufficiently large configuration spaces, including up to three-particle-three-hole ($3p-3h$) components. They are the most important prerequisite for quantitative descriptions of the fine and coarse properties of nuclear spectral functions and description of nuclear data with the highest accuracy \cite{Tso04,Tso08,Tso16,Tso17,Tso20,Tso21}. 
Furthermore, the approach was able to predict the nuclear structure properties and dynamics of new modes of nuclear excitation at low energies and, in particular, the \ac{PDR} and its higher multipole extension, the pygmy quadrupole resonance (PQR) \cite{Tso11}. In this respect, the calculations of branching ratios to the ground and low-energy excited states serve as highly sensitive observables for distinguishing simple one-particle-one-hole ($1p-1h$) type configurations, and multiphonon structures \cite{Tso19,Tso21}. 

With this project, we intend to apply the \ac{EDF} - \ac{QPM}  approach in studies of electric and magnetic low-energy excitations with complex structure, pygmy and giant resonances in atomic nuclei with masses around and below $A$ = 60. Calculations of transition strengths and branching ratios of \ac{PDR} and \ac{GDR} to the lowest-lying excited states and the ground state can provide insight into the mechanism of decay of these excitation modes \cite{Tso21}. 
The \ac{PDR} can contribute to the neutron capture cross sections of nuclear reactions of $s$- and $r$-processes of nucleosynthesis in astrophysics \cite{Tso13,Tso15,Tso17}.
In addition, multiparticle-multihole excitations related to nuclear polarization can cause redistribution and fragmentation of the low-energy $1p-1h$ strength, which can significantly affect radiation capture cross sections \cite{Tso15,Tso16}. In this regard, \ac{EDF} - \ac{QPM}   spectral distributions can be implemented in nuclear reaction codes for evaluating and predicting nuclear reaction rates of astrophysical importance.

\subsection{AMD}
The electric dipole states of light nuclei will be studied by means of the \ac{AMD} model~\cite{kimura2016,taniguchi2004} combined with the shifted-basis method~\cite{kimura2017structure} or the \ac{REM}~\cite{imai2019real}. The decay branches ($n$, $p$, $\gamma$, and $\alpha$ decays) will also be studied within the same framework using the Laplace expansion method~\cite{chiba2017}.

The \ac{AMD} framework uses the Slater determinant wave function of nucleon wave packets to describe the structure and response of atomic nuclei. To incorporate with the particle-hole states which are relevant to the electric dipole states, the shifted-basis method or recently developed \ac{REM} will be combined with \ac{AMD}. The shifted-basis method was already applied to $^{26}$Ne and successfully described the \ac{PDR}, \ac{GDR} and their decay patterns~\cite{imai2019real}. 

The advantages of the model are as follows. 
(1) \ac{AMD} wave function can describe both the collective states and cluster states, such as $\alpha$ clustering, which is essential to estimate the $\alpha$-decay branch. 
(2) The rotational symmetry is restored by the angular momentum projection. This enables us to describe not only the vibrational modes but also the rotational modes and the rotation-vibration coupling. 
(3) The \ac{REM} handles the many-particle-many-hole states, which enables the investigation of the many-body correlations beyond PRA approximation. 
(4) Using the Laplace expansion model, the decay branches can be estimated reasonably.

The following issues are planned to be resolved.
(1) The effective interaction that reasonably describes the structure and response of the light nuclei needs to be developed since the current Skyrme density functionals do not work well for very light nuclei. 
(2) This project will be the first application of \ac{REM} to \ac{AMD}. Computer code development and benchmarking are required. 

\subsection{Large-scale shell-model (LSSM) calculations}


Shell-model calculations are quite successful in systematically reproducing low-lying states and can be a promising option for describing photo-nuclear reactions. 
The photo-nuclear reactions introduce parity change in the excitation from the ground state, demanding much larger valence shells. The minimum model space to satisfy the $E1$ sum is the $0\hbar\omega$ space for the ground state and the $1\hbar\omega$ space for the $E1$-excited states. Namely, when the Fermi surface is located in the major shell having the harmonic-oscillator quantum number $\mathcal{N}_0\hbar\omega$, one has to take three major shells with $(\mathcal{N}_0-1)\hbar\omega$, $\mathcal{N}_0\hbar\omega$, and $(\mathcal{N}_0+1)\hbar\omega$ as the valence shell, and consider all the possible basis states whose harmonic-oscillator quantum numbers are $1\hbar\omega$ larger than the lowest. Such calculations are usually referred to as the $1\hbar\omega$ calculations. 

\begin{figure}
\begin{center}
\resizebox{0.4\textwidth}{!}{%
  \includegraphics{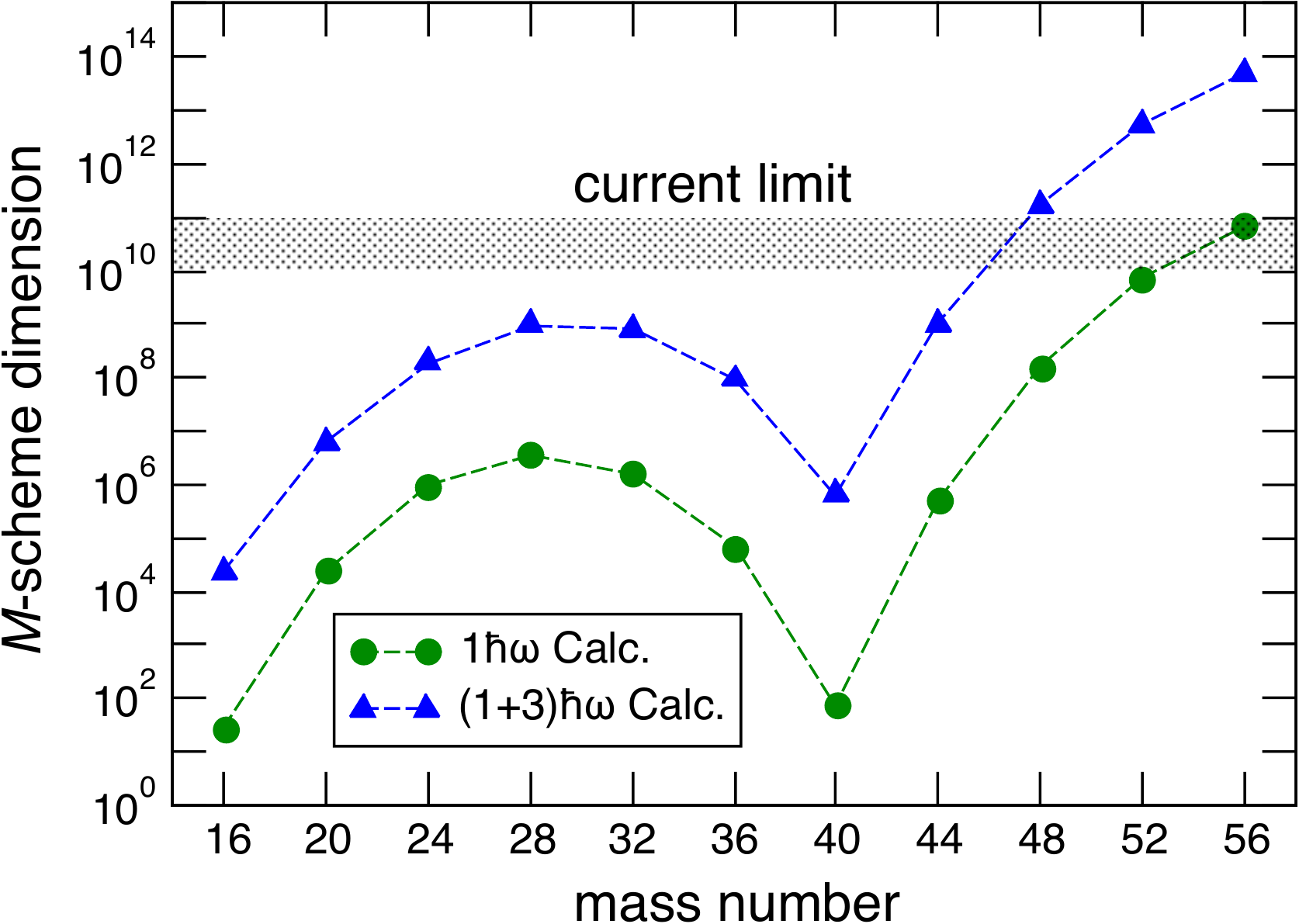}
}
\caption{$M$-scheme dimensions for the $1^-$ states of $N=Z$ even-even nuclei. 
}
\label{fig:dim_sm}  
\end{center} 
\end{figure}

Figure~\ref{fig:dim_sm} presents the $M$-scheme basis dimensions for the $1^-$ states of $N=Z$ even-even nuclei ranging from $^{16}$O to $^{56}$Ni. The circles denote those of the $1\hbar\omega$ calculations. For the nuclei with $16<A<40$, the valence shell consists of the $\mathcal{N}_0=1,2,3$ major shells, and the $M$-scheme basis dimension peaks at the middle of the $sd$ shell, $^{28}$Si. For the nuclei with $A>40$, the valence shell changes to the $\mathcal{N}_0=2,3,4$ major shells. The dimension steeply increases with the mass number than that for $16<A<40$, and reaches about $10^{11}$ at $^{56}$Ni. Since recent large-scale shell-model codes can handle $10^{10}$-$10^{11}$ $M$-scheme dimensions, we will be able to carry out $1\hbar\omega$ calculations for almost all the nuclei of interest using the KSHELL code \cite{KSHELL}. 

The next leading order to be included as the basis states are the $3\hbar\omega$ ones. These states consist of (i) one-nucleon excitation to the $3\hbar\omega$ larger shell and (ii) three-nucleon excitations to the $1\hbar\omega$ larger shells. Since the basis states of (i) require much more valence shells beyond the capability of the current version of KSHELL, we can include only the basis states of (ii) at this moment. We have confirmed that the missing $E1$ strengths due to the lack of (i) are negligible and refer to such approximate calculations as $(1+3)\hbar\omega$ calculations. As shown in Fig.~\ref{fig:dim_sm}, the $(1+3)\hbar\omega$ basis dimensions are $\sim 3$ orders larger than the $1\hbar\omega$ ones. The dimensions for the $sd$-shell nuclei (i.e., $16<A<40$ ones) are, however, still below our limit. 
The $1\hbar\omega$ and $(1+3)\hbar\omega$ calculations are compared in \cite{utsuno2015} for $^{48}$Ca. While the $E1$ strengths below the \ac{GDR} peak are almost the same, those above the \ac{GDR} peak are reduced in the $(1+3)\hbar\omega$ calculation and appear closer to the experimental data probably due to improved correlations. Keeping that in mind, we consider that the best strategy for the \ac{PANDORA} project is to carry out   $(1+3)\hbar\omega$ calculations for the $sd$-shell and lighter $pf$-shell  nuclei and to perform $1\hbar\omega$ calculations with empirical corrections for higher excitation energies included.

Furthermore, numerical calculations derived from different LSSM codes and interaction schemes can be compared and used for theoretical predictions of nuclear excitations in light nuclei of interest for the \ac{PANDORA} project.

\subsection{{\em Ab initio} no-core shell model calculation}


Recently {\em ab initio} studies of the giant dipole resonance have been achieved in light nuclei, such as 
$^{4}$He \cite{horiuchi2012}, $^{10}$B \cite{kruse2019no} and $^{16}$O \cite{soniaO16}.
In these {\em ab initio} approaches, one of the most challenging obstacles is to treat the hard-core nature of nuclear forces, which often demands prohibitively large model space in no-core shell-model calculations.

In Ref.~\cite{kruse2019no}, no-core shell-model calculations successfully described the giant dipole resonance by employing the modern chiral N$^3$LO interaction.
In this work, the similarity-renormalization-group (SRG) evolution was applied to the chiral N$^3$LO interaction to soften the hard-core nature of the nuclear force, and it was demonstrated that the giant dipole resonance obtained by the no-core shell model approach converges with a relatively small model space. 
In this project, we will perform no-core shell model calculations to obtain the giant dipole resonances of light nuclei using another softened interaction, such as the Daejeon 16 interaction \cite{daejeon16}, employing the KSHELL code \cite{KSHELL}.
In order to treat large 
active shells required in no-core shell-model calculations, we may have to develop the KSHELL code further. Moreover, it is desirable to construct an effective $E1$ operator to which the bare $E1$ operator is renormalized in the model space.
In parallel, we will perform no-core shell model calculations employing conventional phenomenological interactions such as the WBT interaction \cite{wbt} for simplicity and for comparison with the {\em ab initio} approaches.


\section{Reaction calculations} 



To estimate photoreaction cross sections of interest in the \ac{PANDORA} project, the different reaction mechanisms at stake need to be simulated. 
One of the modern reaction codes called \talys\ \cite{Koning12} includes many state-of-the-art nuclear models to cover all main reaction mechanisms encountered in light particle-induced nuclear reactions. \talys\  provides a complete description of all reaction channels and observables. The code includes photon, neutron, proton, deuteron, triton, $^3$He, and $\alpha$-particles as both projectiles and ejectiles, and single-particle  as well as multi-particle emissions and fission. All experimental information on nuclear masses, deformation, and low-lying states spectra is considered, whenever available \cite{Capote09}. If not, various local and global input models have been incorporated to represent the nuclear structure properties, optical potentials, level densities, $\gamma$-ray strengths, and fission properties. The \talys\ code was designed to calculate total and partial cross sections, residual and isomer production cross sections, discrete and continuum $\gamma$-ray production cross sections, energy spectra, angular distributions, double-differential spectra,  as well as recoil cross sections. \talys\ also estimates thermonuclear reaction rates of particular relevance to astrophysics \cite{Goriely08a}. 

\begin{figure}
\begin{center}
\resizebox{0.5\textwidth}{!}{%
  \includegraphics{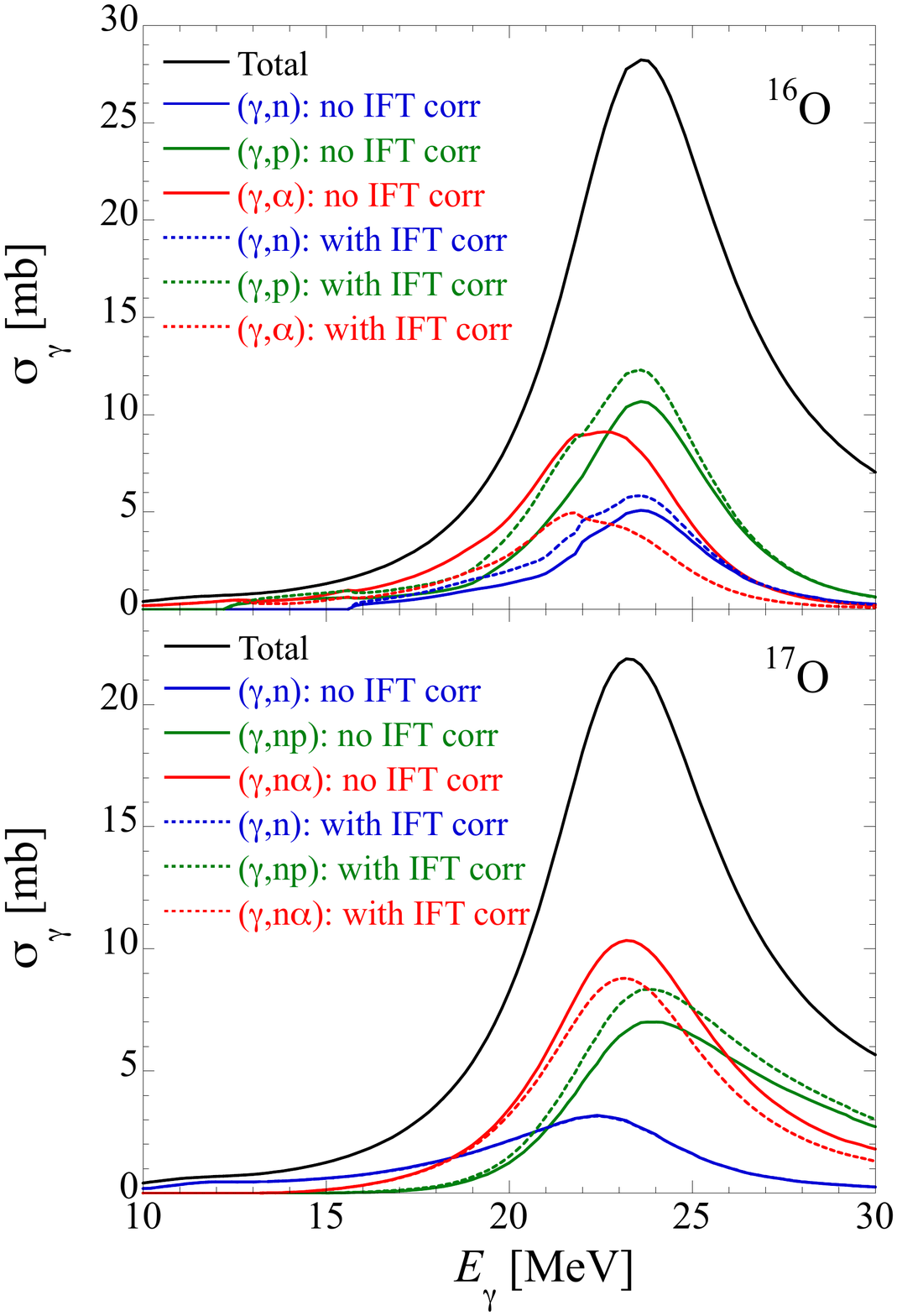}
}
\caption{(Color online) {\sf TALYS} prediction of the dominant photoreaction cross sections on $^{16}$O and $^{17}$O with or without including the IFT corrections. }
\label{fig:talys}
\end{center}
\end{figure}

Of particular relevance to the \ac{PANDORA} project is the treatment of the photon strength function. Different models of the dipole strength function are available in \talys\, in particular, the Simple Modified Lorentzian model (SMLO) \cite{Goriely18a}, or the Gogny D1M+\ac{QRPA} model \cite{Goriely18b}, but experimental strength can also be directly introduced. In addition, for the present project, two recent improvements have been brought to the \talys\ code, namely {\it (i)} an updated description of the nuclear level density of light nuclei within the constant-temperature model \cite{Koning08} where the model parameters have been adjusted individually to all light targets of interest in the present project, and {\it (ii)} the account of isospin forbidden transitions (IFT) both in the single and multiple particle emission channels. For the latter, a phenomenological correction reflecting the hindrance of dipole emission in self-conjugate nuclei is introduced, as detailed in Ref.~\cite{Holmes76}. Including IFT corrections can lead to rather different predictions of the photoemission cross sections in targets like $^{16}$O or $^{17}$O, as illustrated in Fig.~\ref{fig:talys}.

Based on the statistical model of Hauser-Feshbach, \talys\ is known to be a very successful code for medium-mass and heavy target nuclei. Such a model makes the fundamental assumption that the capture process takes place with the intermediary formation of a compound nucleus in thermodynamic equilibrium. The energy of the incident particle is then shared more or less uniformly by all the nucleons before releasing the energy by particle emission or $\gamma$-de-excitation. The formation of a compound nucleus is usually justified by assuming that the level density in the compound nucleus at the projectile incident energy is large enough to ensure an average statistical continuum superposition of available resonances. However, when the number of available states in the compound system is relatively small, as for light targets, the validity of the Hauser-Feshbach predictions has to be questioned. For this reason, {\it a priori}, the statistical model is not well suited for describing the reaction mechanisms taking place with the light species investigated in the current project. However, it appears to provide rather fair rates that can be used as a first guess for sensitivity analysis and further adjusted in specific cases, as shown in Ref.~\cite{Coc12}. In particular, as illustrated in Fig.~\ref{fig:talys}, the IFT corrections can be used to tune photoreaction cross sections close to the $N=Z$ line.

%
%

\section{UHECR physics} 
\label{sec:UHECR}

\subsection{Observations and open questions}

\begin{figure}
\begin{center}
\resizebox{0.45\textwidth}{!}{%
  \includegraphics{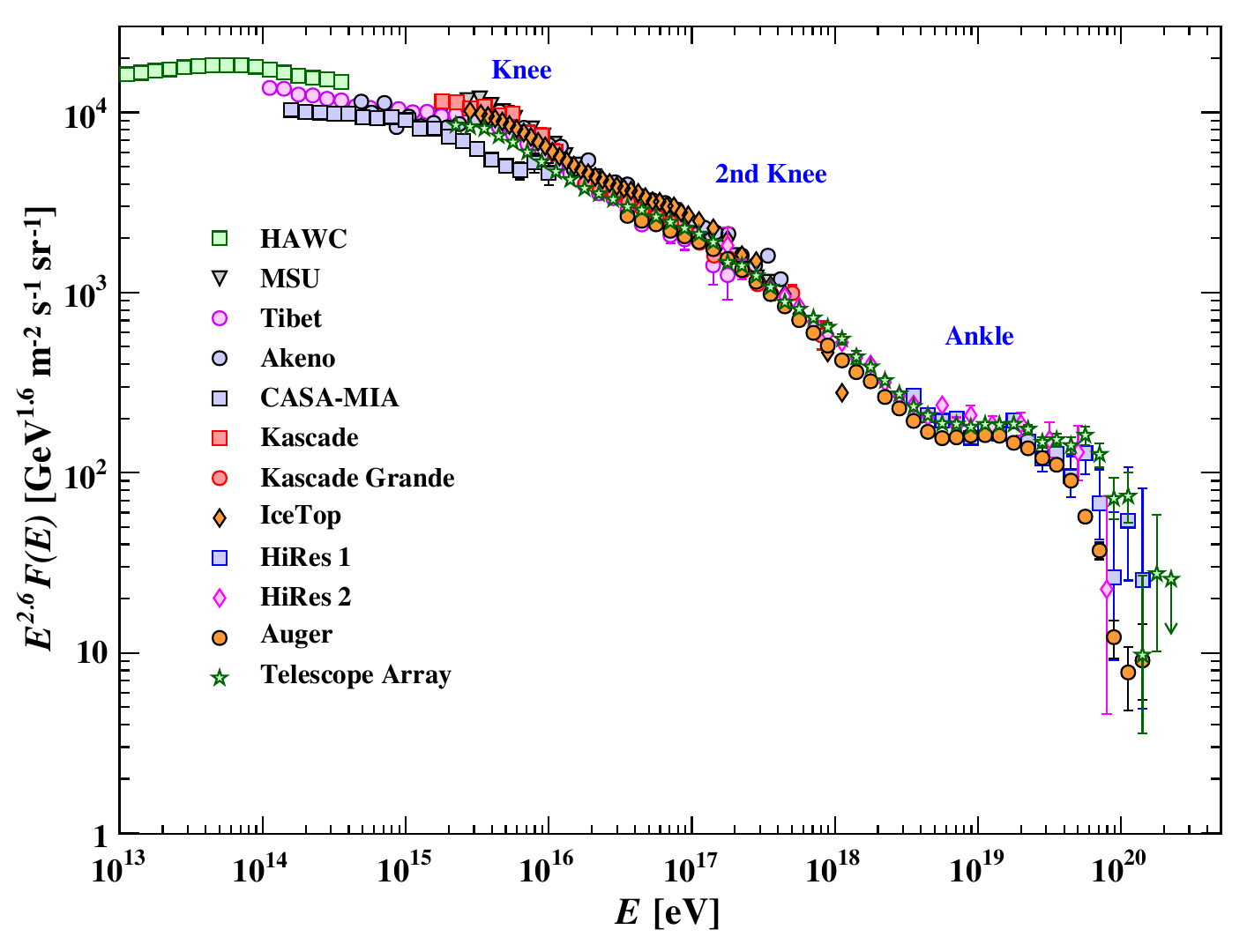}
}
\caption{
The observed cosmic-ray flux distribution as a function of the (total) energy~\cite{PDG2020} above 10 TeV, including data from major \ac{UHECR} observatories such as Pierre Auger (the orange circles) and Telescope Array (the green stars). A cut-off is visible above a few $10^{19}$ eV.
}
\label{fig:CR}
\end{center}
\end{figure}

\begin{figure}
\begin{center}
\resizebox{0.45\textwidth}{!}{%
  \includegraphics{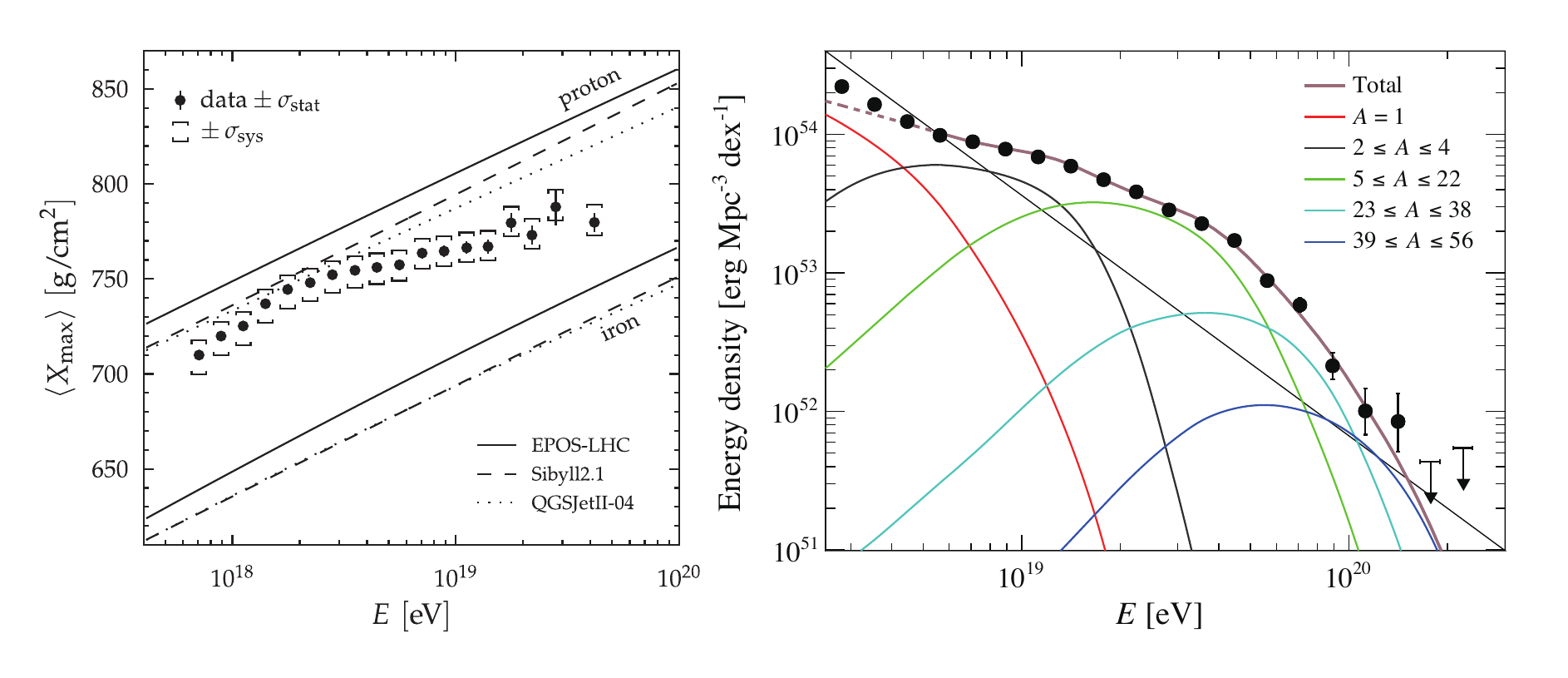}
}
\caption{
The atmospheric depth of the maximum air shower, $<X_{\rm max}>$, as a function of the \ac{UHECR} energy from the Pierre Auger Observatory in comparison with three model predictions for the cases of primary \ac{UHECR} particles of protons and irons (left panel)~\cite{aab14a}. The observed \ac{UHECR} flux distribution above $10^{18}$ eV from the Pierre Auger Observatory compared with an example mass-composition analysis (right panel)~\cite{aab2020}.
}
\label{fig:UHECR_Spectra}
\end{center}
\end{figure}

After more than fifty years of experimental efforts, the origin of \ac{UHECR}s, cosmic-rays above $10^{18}$ eV remains a mystery. 
Understanding the production of these cosmic rays, considered the most energetic particles in the universe, is one of the most intense research fields of high-energy astrophysics. 
High-resolution and high statistics measurements of the \ac{UHECR} spectrum, composition and arrival direction recently became possible through experiments like AGASA~\cite{nag92}, HiRes~\cite{abb08}, the Pierre Auger Observatory~\cite{abr04b} and Telescope Array (TA)~\cite{tok11}, based on the detection of giant air showers triggered by the interaction of \ac{UHECR}s in the Earth atmosphere. 
The evidence of a suppression of the \ac{UHECR} flux above $3-5\times10^{19}$ eV observed by HiRes~\cite{sok07}, the Pierre Auger Observatory~\cite{abr08}, and TA~\cite{abu13,abb16} (see Fig.~\ref{fig:CR}) was one of the most anticipated observation at the highest energies. 
Composition analyses at the Pierre Auger observatory strongly favor a mixed composition with an evolution toward heavier elements (but still most probably not heavier than Fe nuclei) above a few $10^{18}$ eV~\cite{abr10,aab14a,aab14b} (Fig.~\ref{fig:UHECR_Spectra}). This quite unexpected trend shows the central importance of complex nuclei ($A=2$ and heavier) in \ac{UHECR} physics. In addition, the presence of these complex nuclei shows that \ac{UHECR}s are accelerated in astrophysical sources rather than directly produced by exotic high-energy particle physics phenomena.
Anisotropies in the arrival directions of \ac{UHECR}s are additional critical pieces of information to constrain their origin. Recent data provided the first clear evidence (above the $5\sigma$ significance threshold) of a large-scale anisotropy (a dipole modulation) in the distribution of the arrival direction of the highest energy events~\cite{aab17a}. Moreover, Auger and TA data show hints of anisotropies at smaller angular scales ($\sim 20-25^\circ$) with claimed excesses in the direction of Centaurus~\cite{aab15} and Ursa Major~\cite{tin15}.
Despite this evidence for anisotropies, the astrophysical production sites of \ac{UHECR}s remain challenging to trace. This is probably a consequence of significant angular deflections experienced by \ac{UHECR}s, due to cosmic magnetic fields, during their journey from their sources to the Earth. Nevertheless, these anisotropy observations seem to support the general idea that \ac{UHECR}s are of extragalactic origin (as already accepted, almost universally, based on simple theoretical considerations).

\subsection{Extragalactic propagation of UHECRs}

Due to their likely extragalactic origin, the \ac{UHECR} spectrum and composition measured on Earth have to be shaped by the effect of the propagation of the particles in the extragalactic medium. During their journey from the sources (expected to be distributed somewhat like ordinary matter throughout to entire universe) to the Earth, the injected cosmic-ray spectrum and the mass composition are modified by the interactions of \ac{UHECR}s with background photons (resulting in energy and mass losses) and cosmic magnetic fields (resulting in deflections). Detailed modeling of the extragalactic propagation of \ac{UHECR}s is then a necessary ingredient for the astrophysical interpretation of the data. One key feature induced by \ac{UHECR}s extragalactic propagation is the prediction of a cut-off in the observed spectrum above a few $10^{19}$ eV related to interactions of UHE protons or nuclei with photons from the \ac{CMB} as well as with infrared, optical and ultra-violet extragalactic backgrounds (hereafter IR/Opt/UV). This prediction~\cite{gre66,zat66} of the so-called GZK cut-off (named after the authors of the original studies Greisen, Zatsepin, and Kuzmin) was made in 1966, closely following the discovery of the \ac{CMB} and seem to be consistent with the above-mentioned spectral measurements. 


\begin{figure}
\begin{center}
\resizebox{0.5\textwidth}{!}{%
  \includegraphics{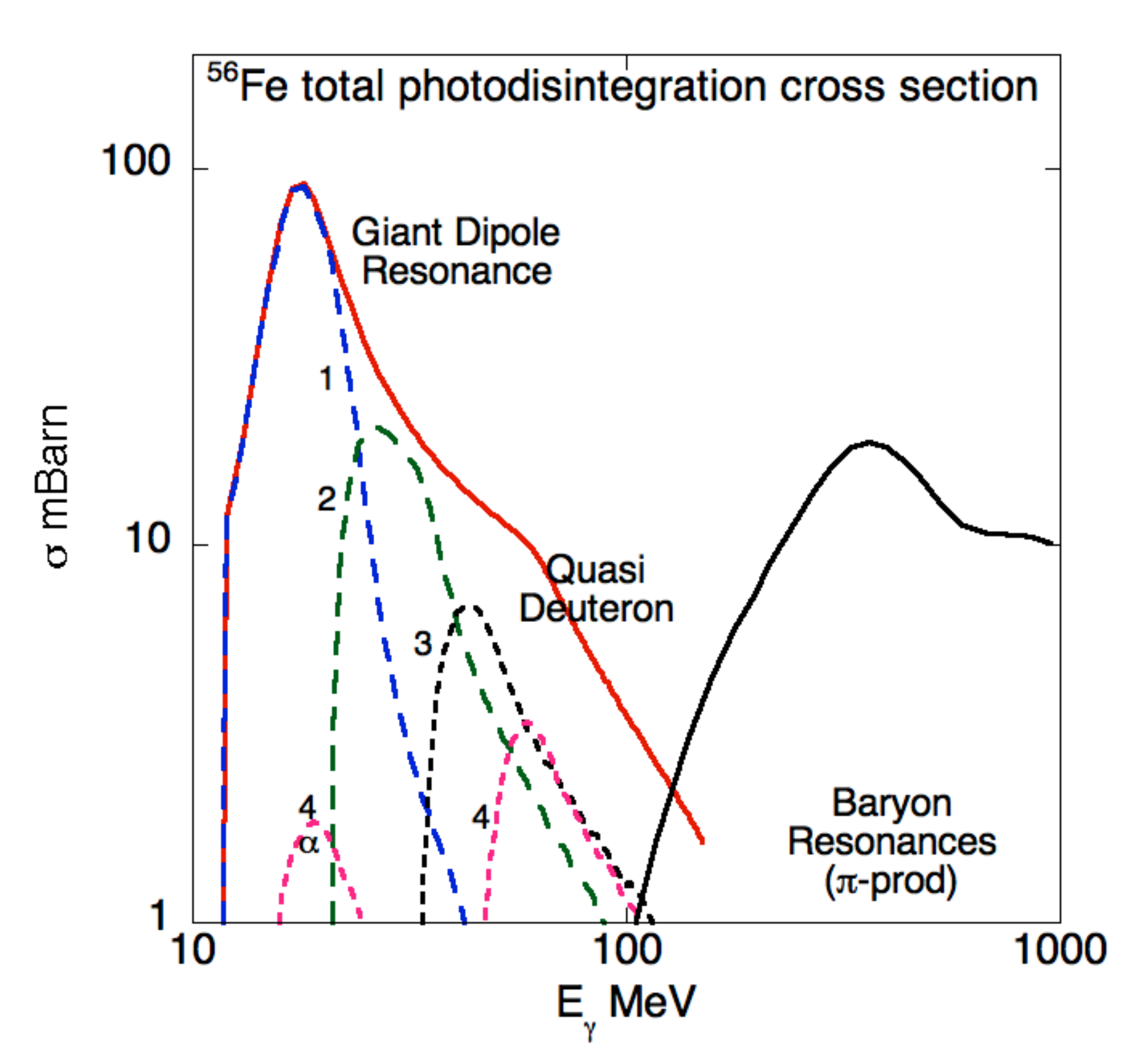}
  \includegraphics{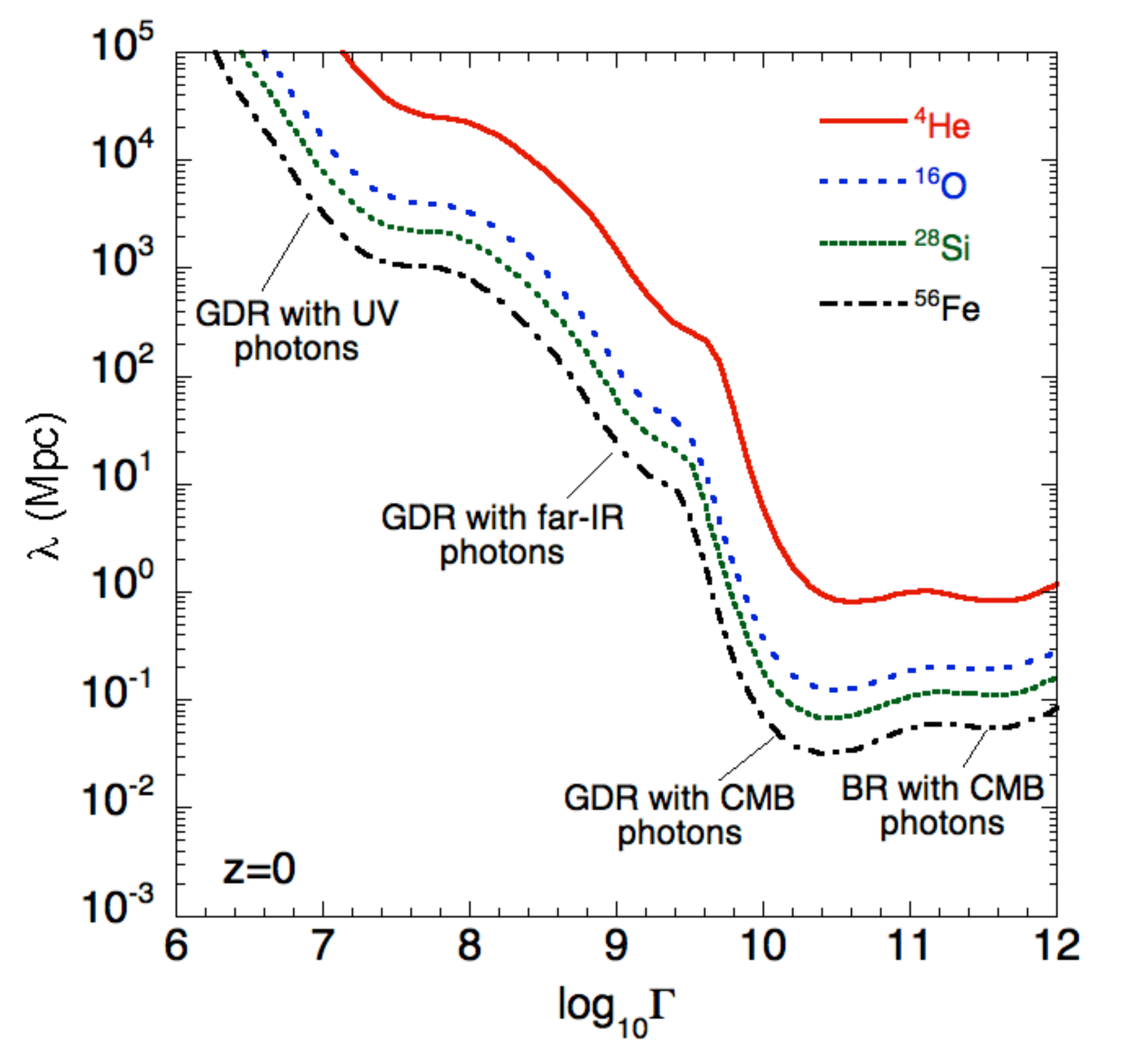}
}
\caption{
Left: Energy evolution of the photodisintegration cross-section for $\rm^{56}Fe$, the contributions of the \ac{IVGDR}, quasi-deuteron (QD) and baryon resonances (BR) are shown as well as the contribution of different nucleon multiplicities (for \ac{GDR} and QD). The cross-sections in the \ac{GDR} and QD regimes were obtained using the \talys\ code. Right: Total photodisintegration mean free path for various species as a function of the Lorentz factor (see labels) at a redshift $z$ = 0 (adapted from \cite{all12}).}

\label{fig:UHECR_Propagation}
\end{center}
\end{figure}

\begin{figure}
\begin{center}
\resizebox{0.45\textwidth}{!}{%
  \includegraphics{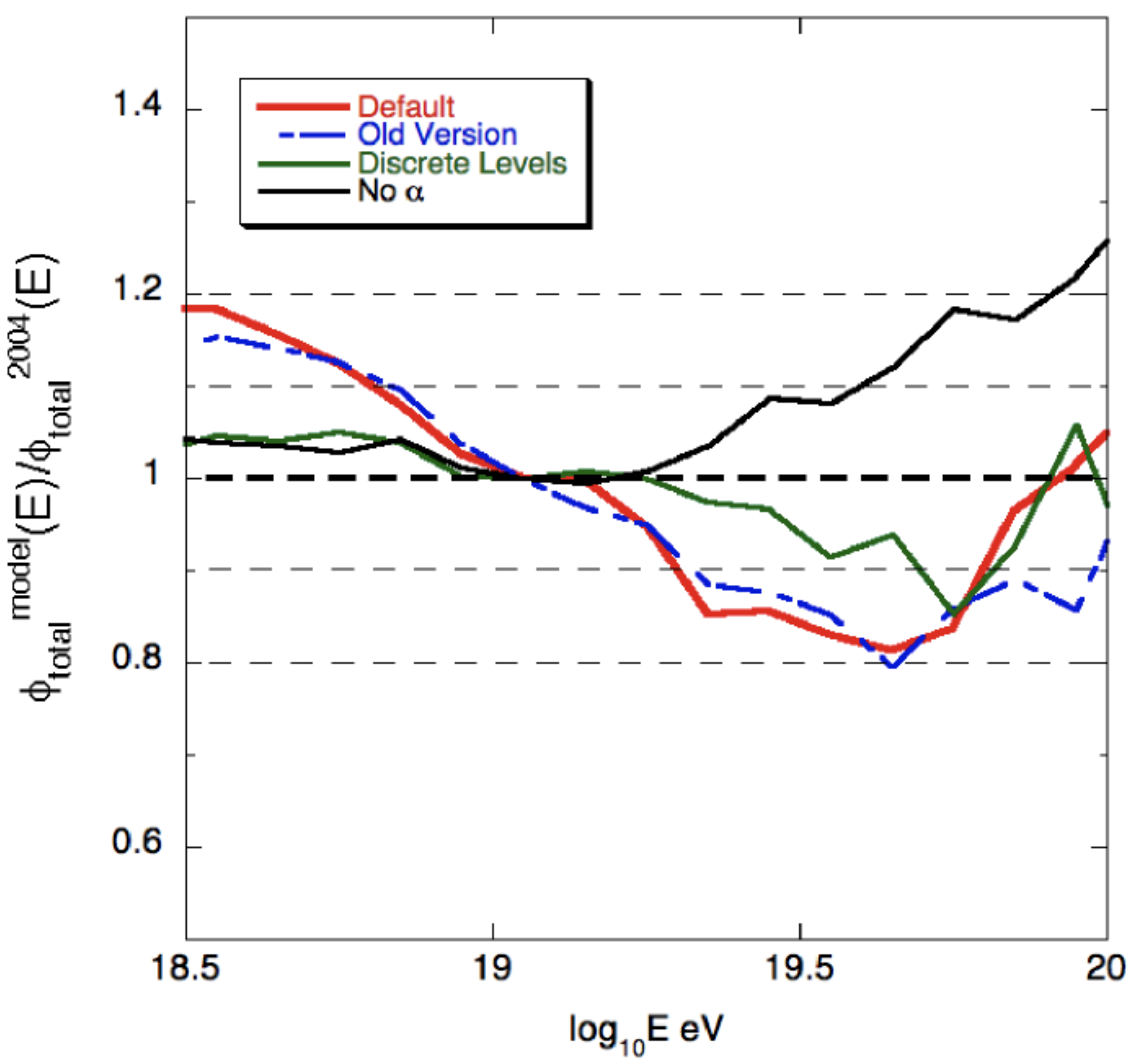}
}
\caption{
Illustration of the effect of the $\alpha$ decay channel on the expected flux after propagation in the intergalactic space, assuming a uniform distribution of sources in the universe, for a few different theoretical assumptions on the \ac{GDR} cross-sections corresponding to various \talys\ settings.
The graph shows the relative difference between the \ac{UHECR} fluxes on Earth predicted, as a function of energy, for the various assumptions. The flux obtained for the cross sections of \cite{kha05} are used as a reference. 
}
\label{fig:UHECR_alpha_decay_dep}
\end{center}
\end{figure}

Only photo-interactions are relevant in the extragalactic medium, except maybe in the immediate vicinity of the \ac{UHECR}s acceleration site.
The principle of the photo-interactions suffered by \ac{UHECR} protons and nuclei during their intergalactic journey, with photon backgrounds of energies ranging from $10^{-3}$ eV (\ac{CMB}) to a few eV (UV) in the Earth frame, is relatively simple. Cosmic-ray protons or nuclei with a Lorentz factor large enough would see photons of these low-energy backgrounds as multi-MeV photons in their proper reference frame, that is, above threshold energies for the $e^+e^-$ pair production or the pion production for protons or nucleon separation for complex nuclei. Whenever happening, each of these interactions results in energy losses for the incoming ultra-relativistic cosmic ray either by decelerating or losing nucleons. As a result, complex nuclei composing \ac{UHECR}s suffer mainly from \ac{IVGDR}, hereafter more shortly \ac{GDR} excitations with extra-galactic photons for Lorentz factors above $\sim10^{8.5}$ with far-infrared photons and $\sim10^{9.5}$ with \ac{CMB} photons (see Fig.~\ref{fig:UHECR_Propagation}). This results in nucleon losses and modification in the \ac{UHECR} composition with respect to that produced in the unknown sources. Note that \ac{GDR} is the most important photodisintegration process for \ac{UHECR} nuclei since it has the largest cross-section and the lowest energy threshold, it thus has a dominant contribution to the \ac{UHECR} nuclei with photodisintegration mean free path for Lorentz factor ranging from $\sim 10^7$ to $10^{11}$ (see Fig.~\ref{fig:UHECR_Propagation}).

The most significant early study of \ac{UHECR} nuclei propagation was certainly made by Puget, Stecker and Bredekamp (PSB)~\cite{pug76} who treated, in great detail, the propagation of nuclei ($A\le56$) and provided simple parametrizations based on the available data for the \ac{GDR} and quasi-deuteron (QD) excitation cross sections. More recently, Khan et al.~\cite{kha05} proposed new estimates of the \ac{GDR} cross-section, based on a theoretical calculation using the \talys\ nuclear reaction code~\cite{kon04} showing a better agreement with the available data than previous parametrizations from PSB~\cite{pug76}. The improvement in the \ac{UHECR} estimates brought by these new calculations rely not only on the use of microscopic cross sections but also on the implementation of a photodisintegration network involving many nuclei in contrast to what was done previously were only one given nucleus was considered for each $A$ below 56. 

In recent years the interest in the propagation of \ac{UHECR} nuclei has significantly grown (see Ref.~\cite{all12} for a recent review and references therein). This is all the more true since the recent composition analyses of the Pierre Auger Observatory now firmly indicate a significant (and increasing with energy) contribution of nuclei at the highest energies. The availability of \ac{GDR} cross section models as well constrained and as realistic as possible is, of course, critical to obtain the most information out of extragalactic  \ac{UHECR} propagation studies. Some predictions of nuclear reaction codes such as \talys\ (see previous sections), for instance, the importance of the $\alpha$-decay reaction channels for low mass nuclei, depend on the chosen settings of the nuclear reaction code and are, on the other hand, not currently well constrained by nuclear physics measurements. The latitude left by the lack of experimental constraints, together with the dependence on the nuclear reaction code settings, can lead to a sizeable dispersion of the \ac{UHECR} spectrum predicted after accounting for the \ac{UHECR} extragalactic propagation for a given source spectrum and composition model as illustrated in Fig.~\ref{fig:UHECR_alpha_decay_dep}. 
The dispersion of \ac{UHECR} spectrum was also evaluated in Ref.~\cite{Kido22} using the \ac{RPA}-\ac{EDF} model~\cite{inakura2009,inakura2011} with several Skyrme-\ac{EDF} parameters. 

An exhaustive campaign of measurements of \ac{GDR} cross sections for various nuclei and reaction channels as proposed in the framework of the \ac{PANDORA} could contribute to significantly improving the description of nuclear reactions relevant to the propagation of \ac{UHECR}s. 
Likewise, these improvements would also benefit the modeling of \ac{UHECR} acceleration in potential \ac{UHECR} astrophysical sources, including, for instance, 
active galactic nuclei~\cite{1987ApJ...322..643B,1993A&A...272..161R,2008MNRAS.388L..59G,2008MNRAS.391.1100F,2011APh....34..749T,2019MNRAS.482.4303M}, 
gamma-ray bursts~\cite{1995PhRvL..75..386W,1995ApJ...452L...1W,2003ApJ...592..378V,2011MNRAS.415.2495M}, 
star-burst galaxies~\cite{1999PhRvD..60j3001A,2018PhRvD..97f3010A,2022MNRAS.511.1336P}, 
galaxy clusters~\cite{1996ApJ...456..422K,1998PhRvD..59b3001B,2019SciA....5.8227K}, 
or tidal disruption events~\cite{2017PhRvD..96f3007Z},
where photo interactions could also play an important role.

\section{Applications}
\label{sec:applications}
\subsection{Astrophysical applications beyond UHECRs}

In addition to their relevance to a proper understanding of the \ac{UHECR} composition and energy spectrum as detected on Earth (Sect.~\ref{sec:UHECR}), photoreactions also play an important role in stellar evolution and nucleosynthesis applications \cite{Arnould20}. More specifically, at high temperatures (typically in excess of about 10$^9$K), nuclei may be subject to ($\gamma$,n), ($\gamma$,p) or ($\gamma$,$\alpha$) photodisintegrations. These transformations become more and more important as the evolution of a star proceeds beyond C-burning in the non-explosive history of massive stars. The first major stage of photodisintegrations is the so-called Ne-burning episode governed by the $^{20}$Ne($\gamma,\alpha$)$^{16}$O reaction which also produces a significant amount of $\alpha$-particles that are recaptured and responsible for the production of elements heavier than Ne. Photodisintegrations culminate at the Si-burning phase terminating with a nuclear statistical equilibrium. Explosive situations favor photodisintegrations due to the higher temperatures reached in these conditions. In fact, some can already occur during the various hot modes of H-burning \cite{Iliadis15} and are essential ingredients of the p- and r-processes of nucleosynthesis \cite{Arnould03,Arnould07}.

The photodisintegration rates in a photon bath obeying a Plank distribution law that applies in stellar interiors are briefly discussed in Ref.~\cite{Iliadis15}. In view of the difficulties of deriving photodisintegration rates through direct approaches, they are traditionally estimated based on the detailed balance theorem applied to the reverse radiative captures of nucleons or $\alpha$-particles ({\it e.g.} \cite{Holmes76,Iliadis15}). However, photoreaction experiments can be of great interest to constrain the reverse cross sections, especially when dealing with an unstable target in the radiative capture that cannot be measured in the laboratory. For example, the photodisintegration of $^9$Be through the $1/2^+$ state near neutron threshold \cite{Utsunomiya15} helped to constrain the important three-body inverse reaction ($\alpha \alpha$n) of relevance during the nucleosynthesis in the neutron-rich $\nu$-driven wind of type-II supernovae.

\subsection{Application to other fields}
Precise knowledge of the $E1$ transition strength distribution and the decay processes is essential for various applications related to
$\gamma$-ray irradiation or photo-excitation.
The excitation-energy distribution of the $E1$ transition strength determines the photo-absorption cross sections and, thus, the reaction rate under the $\gamma$-ray irradiation.
Reliable predictions of the decay processes are indispensable for simulating radiation effects, {\it e.g.} the number of nuclei produced after photo-excitation, energy dissipation by the emitted charged particles, amount of the produced neutron radiations and the activation by the neutrons, production of radioactivities, and emission of characteristic $\gamma$-rays used for elemental or isotopic analysis.
Below applications of the photo-nuclear reaction data and predictions of light nuclei are briefly summarized.

The photo-nuclear reaction data are essential inputs for the simulation of radiation shielding design, radiation transport analysis, and estimation of radioactivity for decommissioning
at industrial, scientific, and medical facilities.

The method of photon activation analysis (PAA) has been developed for identifying non-destructively the nuclides in a sample material by detecting the induced radioactivity after exposure to high-energy photons~\cite{segebade2017principles}.
PAA is more suitable than neutron activation analysis (NAA) for the case of detecting light elements such as Be, C, N, O, F, Si, P, and Ni~\cite{ibrahim2020measurement}.
Super-intense Laser-driven photon activation analysis is proposed by using photons produced by high-power laser irradiation to solid material~\cite{mirani2021superintense}.
Gamma-ray imaging by combining the techniques of nuclear resonance fluorescence (NRF) spectroscopy with computer tomography (CT) is developed for non-destructively investigating the distribution of specific isotopes in material~\cite{toyokawa2011two}.

Inspection of fissile material by applying the photo-nuclear reactions is discussed in several articles for international safeguard and security~\cite{johnson2010applications,jones2007high,verbitskii2011applying}. 
Although the photo-neutron response of fissile heavy elements is out of the scope of the \ac{PANDORA} project, the photo-neutron response of surrounding materials, for example the $^{18}{\rm O}(\gamma,n)$ reaction, needs to be evaluated since the reactions produce noise signals for the detection of the target material~\cite{jones2007high}.
Also, photoneutron spectroscopy using monoenergetic $\gamma$-rays for detecting bulk explosives is studied by measuring the spectrum of emitted characteristic neutrons to identify nuclei associated with explosives (H, C, N, O) ~\cite{mcfee2013photoneutron}

Production of medical isotopes by photon irradiation is investigated for e.g. $^{44}{\rm Sc}$, $^{47}{\rm Ca}$ by $(\gamma,n)$, $^{47}{\rm Sc}$ by $(\gamma,p)$, and $^{44}{\rm Ti}$ by by $(\gamma,2n)$ reactions~\cite{habs2011production}.
Precise data on photo-nuclear reactions on the human body's elements are essential for simulating the absorbed doses during radiotherapy. 

Photo-nuclear reactions triggered by lightning discharge was recently observed by terrestrial scintillation $\gamma$ detectors~\cite{enoto2017photonuclear,wada2019downward}.
The ignition process and lightning discharge development are still poorly understood.
The accelerated electrons in the thundercloud emit $\gamma$ rays by Bremsstrahlung. 
The $\gamma$-ray produces $^{13}{\rm N}$ and $^{15}{\rm O}$ nuclei by $^{14}{\rm N}(\gamma,n)$ and $^{16}{\rm O}(\gamma,n)$ reactions in the atmosphere.
The produced nuclei emit positrons with a half-life of several minutes. The annihilation $\gamma$-rays are observed by scintillation $\gamma$-detectors.
Precise knowledge of the photo-nuclear reactions on the relevant nuclides is required to describe the lightning mechanism in thunderclouds before and after the discharge.

\section{Summary and outlook} 
\label{sec:summary}

Photo-nuclear reactions of light nuclei below a mass of $A=60$ are essential for the nuclear structure and reaction studies, particle and nuclear astrophysics, and various applications.
The photo-nuclear reactions are planned to be systematically investigated by the PANDORA project.
The project consists of experimental nuclear physics, theoretical nuclear physics, and particle astrophysics. 
Two modern experimental methods, virtual-photon excitation by proton scattering and real-photo absorption by using high-brilliance $\gamma$-ray beam produced by laser Compton scattering, will be applied to measure the photo-absorption cross sections and the decay branching ratio of each decay channel. 
Several nuclear models, anti-symmetrized molecular dynamics, a few mean-field type models, a large-scale shell model, and {\em ab initio} no-core shell model will be employed to predict the photo-nuclear reactions.
The uncertainties of the model predictions will also be studied.
The results will be implemented in a reaction calculation code \talys\ for applications in various fields.
The photo-disintegration process of ultra-high-energy cosmic rays in inter-galactic propagation using the theoretical predictions and the uncertainty originating from the accuracy of the nuclear model predictions will be studied as a primary motivation of the PANDORA project.

The first beam times of the project's nuclear physics experiments are planned for 2022 at RCNP, and 2023 at iThemba LABS. The first commissioning of the LCS $\gamma$-ray facility at ELI-NP is expected at the end of the year 2023. Systematic data will be obtained in 5-10 years.
In parallel, developments of nuclear models~\cite{Inakura22} and their application to the ~\ac{UHECR} propagation simulation~\cite{Kido22} are ongoing.

\section{Acknowledgements}
This work is supported partly by
Grants-in-Aid for Scientific Research "KAKENHI" in Japan with
grant numbers of JP19H00693 and 22K18715,
Japan-South Africa Bilateral Funding from JSPS with a grant number of JPJSBP 120216502 and from NRF with grant number 132993, 
the National Research Foundation of South Africa through Grants No.
129411, 85509, and 118846,
a grant of the Romanian Ministry of Research, Innovation and Digitization, CNCS - UEFISCDI, project number PN-III-P4-PCE-2021-0595, within PNCDI III,
Pioneering Program of RIKEN for Evolution of Matter in the Universe (r-EMU), and Interdisciplinary Theoretical and Mathematical Sciences program (iTHEMS, https://ithems.riken.jp) at RIKEN,
the Deutsche Forschungsgemeinschaft (DFG, German Research Foundation) - Project-ID 279384907 - SFB1245, the State of Hesse within the Research Cluster ELEMENTS (Project ID 500/10.006),
the grant "Nuclear Photonics" within the LOEWE program,
and EXC-2094 – 390783311 under Germany´s Excellence Strategy.
S.G. acknowledges financial support from the F.R.S.-FNRS (Belgium).

\bibliographystyle{epj}
\bibliography{Pandora}

\acrodef{AMD}{Antisymmetrized Molecular Dynamics}
\acrodef{ALBA}{African Lanthanum Bromide Array}
\acrodef{CAKE}{Coincidence Array for K600 Experiment}
\acrodef{CMB}{Cosmological Microwave Background}
\acrodef{CDFT}{covariant density functional theory}
\acrodef{DFT}{density functional theory}
\acrodef{DSSSD}{double-sided silicon strip detectors}
\acrodef{ELI-NP}{Extreme Light Infrastructure -- Nuclear Physics}
\acrodef{ELIGANT}{ELI Gamma Above Neutron Threshold}
\acrodef{ELIGANT-GN}{ELIGANT Gamma Neutron}
\acrodef{ELIGANT-TN}{ELIGANT Thermal Neutron}
\acrodef{ELISSA}{ELI Silicon Strip Array}
\acrodef{EDF}{Energy-Density Functional}
\acrodef{EOM}{equation of motion}
\acrodef{GDR}{Giant Dipole-Resonance}
\acrodef{GRAF}{Grand RAiden Forward}
\acrodef{IVGDR}{Isovector Giant Dipole Resonance}
\acrodef{LCS}{laser Compton scattering}
\acrodef{MDA}{Multipole Decomposition Analysis}
\acrodef{MWDC}{Multi-Wire Drift Chambers}
\acrodef{NFT}{Nuclear Field Theory}
\acrodef{PANDORA}{Photo-Absorption of Nuclei and Decay Observation for Reactions in Astrophysics}
\acrodef{PDR}{Pygmy Dipole Resonance}
\acrodef{RCNP}{Research Center for Nuclear Physics}
\acrodef{REM}{real-time evolution method}
\acrodef{RNFT}{Relativistic Nuclear Field Theory}
\acrodef{RMF}{relativistic mean field}
\acrodef{RPA}{Random Phase Approximation}
\acrodef{QCD}{quantum chromodynamic}
\acrodef{QHD}{Quantum Hadrodynamics}
\acrodef{QPM}{Quasiparticle-Phonon Model}
\acrodef{qPVC}{quasiparticle - vibration coupling}
\acrodef{QRPA}{Quasiparticle Random Phase Approximation}
\acrodef{RRPA}{Relativistic Random Phase Approximation}
\acrodef{RQRPA}{Relativistic Quasiparticle Random Phase Approximation}
\acrodef{RQTBA}{relativistic quasiparticle time blocking approximation}
\acrodef{SAKRA}{Si Array developed by Kyoto and osaka for Research into Alpha cluster states}
\acrodef{SCRIT}{Self-Confining RI Ion Target}
\acrodef{SLEGS}{Shanghai Laser Electron Gamma Source}
\acrodef{SSC}{Separated Sector Cyclotron}
\acrodef{UHECR}{ultra-high-energy cosmic rays}
\acrodef{VEGA}{Variable Energy Gamma}




\end{document}